\documentclass [acmsmall,screen]{acmart}

\AtBeginDocument{%
  }

\usepackage{amsmath}
\usepackage{algorithmic}
\usepackage{graphicx}
\usepackage{textcomp}
\usepackage{xcolor}
\usepackage{caption}

\usepackage{siunitx}
\usepackage{multirow}
\usepackage{enumitem}
\usepackage{wrapfig}
\usepackage{url}
\usepackage{booktabs}
\usepackage[switch]{lineno} 
\usepackage[most]{tcolorbox} 
\newtcolorbox{answer}{
    colback=gray!20, 
    colframe=white, 
    boxrule=0pt, 
    arc=3pt, 
    left=4pt, 
    right=4pt, 
    top=3pt, 
    bottom=3pt, 
    boxsep=0pt, 
    shadow={0pt}{-0.5pt}{0pt}{black!30} 
}

\setcopyright{acmlicensed}
\copyrightyear{2026}
\acmYear{2026}
\acmDOI{XXXXXXX.XXXXXXX}
\acmConference[FSE '26]{The ACM International Conference on the Foundations of Software Engineering}{Jul 5--9, 2026}{Montreal, Canada}

\acmISBN{978-1-4503-XXXX-X/2018/06}

\begin{document}

\title{Towards Automated Smart Contract Generation: Evaluation, Benchmarking, and Retrieval-Augmented Repair}

\author{Zaoyu Chen}
\authornote{Equal contribution as first authors.}
\email{zaoyu22.chen@connect.polyu.hk}
\orcid{0009-0003-0397-8952}

\author{Haoran Qin}
\authornotemark[1]
\email{23037834r@connect.polyu.hk}

\affiliation{%
  \institution{The Hong Kong Polytechnic University}
  \country{Hong Kong SAR, China}
}

\author{Nuo Chen}
\authornote{Equal contribution as second authors.}
\email{pleviumtan@toki.waseda.jp}

\author{Xiangyu Zhao}
\authornotemark[2]
\email{xiang-yu.zhao@connect.polyu.hk}

\affiliation{%
  \institution{The Hong Kong Polytechnic University}
  \country{Hong Kong SAR, China}
}

\author{Lei Xue}
\affiliation{
  \institution{Sun Yat-sen University}
  \state{Guangdong}
  \country{China}
}
\email{xuelei3@mail.sysu.edu.cn}

\author{Xiapu Luo}
\authornote{Corresponding authors.}
\affiliation{
  \institution{The Hong Kong Polytechnic University}                                
  \country{Hong Kong SAR, China}
}
\email{daniel.xiapu.luo@polyu.edu.hk}

\author{Xiao-Ming Wu}
\authornotemark[3]
\affiliation{
  \institution{The Hong Kong Polytechnic University} 
  \country{Hong Kong SAR, China}                          
}
\email{xiao-ming.wu@polyu.edu.hk}

\renewcommand{\shortauthors}{Chen et al.}

\begin{abstract} 
Smart contracts, predominantly written in Solidity and executed on blockchains like Ethereum, are immutable, making functional correctness paramount: once deployed, bugs and vulnerabilities become permanent. Despite rapid progress in transformer-based code LLMs, existing evaluations of Solidity code completion rely heavily on surface-form metrics (e.g., BLEU, CrystalBLEU) or hand-grading, which poorly correlate with functional correctness. Unlike Python, Solidity lacks large-scale and execution-based benchmarks, hindering systematic assessment and optimization of LLMs for smart contract development.

To bridge this research gap, we introduce \textbf{SolBench}, a comprehensive benchmark and automated testing pipeline for Solidity, designed to emphasize functional correctness via differential fuzzing. SolBench contains \num{28825} functions from \num{7604} contracts collected from Etherscan (genesis–2024), spanning 10 popular domains. We benchmark 14 diverse LLMs (open/closed, 1.3B–671B parameters, general/code-specific, with/without reasoning).
The dominant failure mode is missing crucial details (e.g., type definitions, state variables) in intra-contract context. Providing full-contract context mitigates this and improves code completion accuracy.

However, full-context inference can be prohibitively expensive in practice. Generating outputs with large context windows using state-of-the-art models often incurs significant costs, rendering naive context scaling economically impractical. Crucially, most of a contract is irrelevant to implementing a given function; only a small subset of details is needed. To exploit this, we propose \textbf{Retrieval-Augmented Repair} (RAR), which integrates retrieval into code repair: it uses the executor’s error messages to extract only the most relevant snippets from the full contract. RAR sharply reduces input length for function completion, improving accuracy while significantly cutting computational cost. We further analyze retrieval and code repair strategies within RAR, showing substantial improvements in accuracy and efficiency. SolBench and our RAR framework enable principled evaluation and cost-effective improvement of Solidity code generation. Dataset and code are available at \url{https://github.com/ZaoyuChen/SolBench}.
\end{abstract}



\maketitle

\section{Introduction}
Smart contracts, which are specialized programs executed on blockchains, play a vital role in decentralized applications by autonomously enforcing agreements without the need for intermediaries. These contracts operate on the Ethereum blockchain, and are predominantly written in Solidity. Unlike traditional software development, the creation of blockchain-based smart contracts presents unique challenges due to their immutable nature; once deployed, the code cannot be altered, rendering any errors or vulnerabilities permanent. Consequently, this underscores the importance of prioritizing functional correctness in the development of smart contracts.

\begin{figure}[t]

    \centering
    \includegraphics[width=0.8\columnwidth]{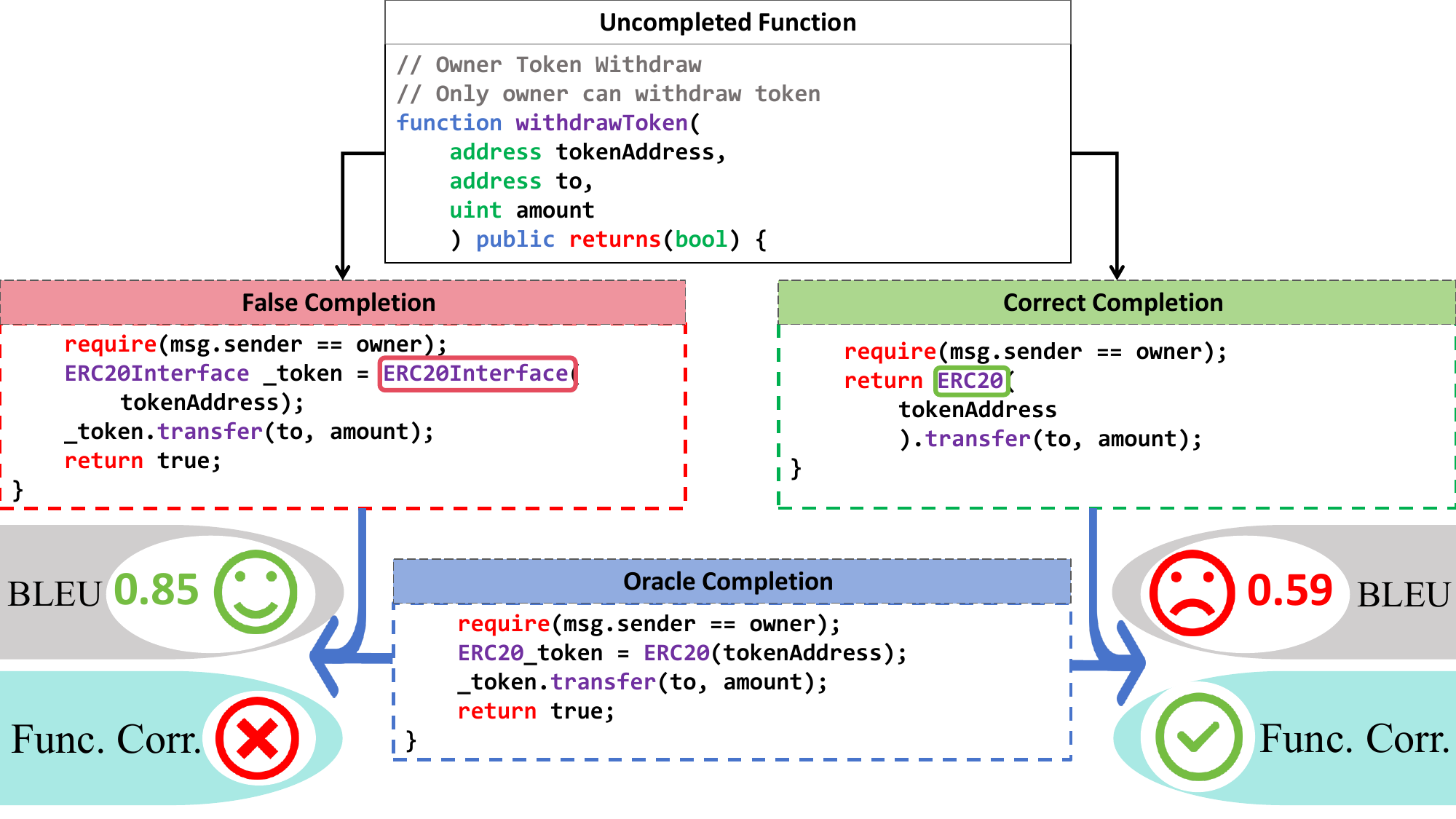} 
    \caption{
A comparison of functional correctness vs. BLEU metric in Solidity code completion, highlighting the importance of evaluating functional correctness.
The model fails to complete the Solidity function (False Prediction) because the comment and signature lacked details. It hallucinates the Interface name as ``\textit{ERC20Interface}''. When given the correct Interface name, ``\textit{ERC20}'', in the prompt, the model identified the Interface name accurately (Correct Prediction). Despite the false prediction receiving a higher BLEU score because of its strong resemblance to the ground truth, our functional correctness metric successfully addresses this problem.
} 
    \label{fig:context}
\end{figure}

Recent advancements in transformer-based Code Large Language Models (LLMs), such as Codex \cite{chen2021evaluating}, CodeGen \cite{nijkamp2022codegen}, StarCoder \cite{li2023starcoder}, and others \cite{roziere2023code,guo2024deepseek}, have generated significant interest in utilizing LLMs to enhance the smart contract development process. 
The current evaluation of smart contract generation predominantly relies on metrics such as BLEU and CrystalBLEU \cite{storhaug2023efficient}, or subjective hand-grading \cite{dade2023optimizing}.
This reliance can misguide model development and selection, as these metrics fail to robustly capture whether the generated Solidity code is functionally correct.
We empirically demonstrate this misalignment by comparing Pass@1 (functional correctness) with BLEU and CrystalBLEU. Pass@1 is more sensitive than BLEU-based scores, and BLEU-based scores show only weak correlation with Pass@1 (Pearson $r \leq 0.41$). This indicates that n-gram overlap poorly reflects actual correctness. In practice, multiple semantically correct implementations may differ in surface form (variable names, control flow, ordering), leading BLEU-style metrics to penalize valid solutions and reward incorrect yet lexically similar outputs.
Figure~\ref{fig:context} also illustrates this inconsistency: a function completed correctly is deemed accurate by functional correctness metrics but receives a low BLEU score.

Languages like Python benefit from mature function-level datasets, test suites, and established benchmarking protocols for functional correctness, such as HumanEval \cite{chen2021evaluating}, APPS \cite{hendrycks2021measuring}, MBPP \cite{austin2021program}, and DS-1000 \cite{lai2023ds}. In contrast, the Solidity ecosystem lacks large-scale, function-level datasets and automated evaluation frameworks. Additionally, Solidity’s unique execution environment requires specialized testing and verification strategies. 
This highlights a research gap in systematically evaluating code LLMs for functional correctness in the context of smart contract generation.

In this work, we present SolBench, a benchmark to address these gaps by providing a dedicated dataset and testing pipeline tailored to Solidity, enabling robust evaluation of smart contract functional correctness. 
The SolBench benchmark comprises \num{28825} functions across \num{7604} contracts sourced from all smart contracts on Etherscan, spanning from the genesis block to the year 2024, and covers 10 popular domains (Table~\ref{tab-contract-type-statistics}).
It uses Differential Fuzzing~\cite{CryticDiffusc2024} to verify whether model-completed functions are functionally equivalent to references, overcoming the limitations of BLEU-style metrics.

We evaluate 14 LLMs (Table~\ref{tab-selected-llms}), chosen to encompass a diverse range of characteristics, including accessibility (open- and closed-source), model scales (1.3B–671B), training focuses (general-purpose vs. code-specific), reasoning capabilities, and release periods. These criteria ensure a comprehensive and representative evaluation.
The experimental results highlight the significant challenges across models, with missing contextual information being the dominant source of errors (Fig.~\ref{fig-error-distr}).
As illustrated in Fig.~\ref{fig-rar-pipline}, function comments and signatures alone are frequently insufficient.
Crucial details (e.g., type definitions, state variables) often reside elsewhere in the same contract, but naively supplying the entire contract is costly: the average context length is 11,654 tokens (Table~\ref{tab-SolBench-statistics}). 
Concretely, running a full-context SolBench evaluation with Claude-Opus-4.1 would cost \$5,286, illustrating that directly scaling context is economically impractical.
Yet most of the contract isn’t needed to implement the current function; we only need the crucial details relevant to that function. Therefore, a precise extraction of only the useful code snippets can complete the function while greatly reducing inference cost.

To address this limitation, we propose a novel Retrieval-Augmented Repair (RAR) framework specifically tailored for Solidity code generation, which reduces inference cost by half while improving accuracy. While Retrieval-Augmented Generation (RAG) techniques have been widely utilized in LLM-based code generation to enhance the relevance and accuracy of generated code by incorporating \emph{external} knowledge or code sources \cite{zhang2023repocoder, shrivastava2023repository, liu2023repobench}, our RAR framework differentiates itself by focusing on searching and retrieving information from the \emph{provided context} of the smart contract rather than external code bases. The RAR framework operates by initially verifying the functional correctness of a function generated by an LLM using an executor. If the code is deemed incorrect, a retriever extracts pertinent snippets from the contract's context based on the executor's traces. Subsequently, the LLM repairs the code utilizing both the retrieved snippets and feedback from the executor. 
By enabling accurate repairs with smaller context windows, RAR reduces inference cost by half while simultaneously improving performance (Sec.~\ref{sec-cost-effective}): with a 1k context window, it matches the performance of a standard model with a 2k window (85.21\% vs. 85.93\% pass rate), and at 2k context, RAR even outperforms a standard 4k context model (89.42\% vs. 88.87\%), as shown in Sec.~\ref{sec-RAR-double}.
Building on this framework, we further investigate the impact of integrating different retrieval strategies with code repair methods on the performance of LLMs in Solidity code completion. 

In summary, our primary contributions include:
\begin{itemize}
    \item 
We present SolBench, a large-scale benchmark and Solidity-specific, fuzz-testing-based evaluation pipeline for function-level functional correctness.
    \item 
We introduce RAR, a Retrieval-Augmented Repair framework that leverages intra-contract context to improve correctness under constrained context windows, achieving the same or better accuracy at roughly half the inference cost.
    \item 
We conduct a comprehensive evaluation of 14 LLMs and various code repair and retrieval techniques within the RAR framework. The results reveal clear differences between models, indicate significant potential for model improvement, and highlight the strong effectiveness of RAR.
\end{itemize}
\section{SolBench: A Large Benchmark for Evaluating Solidity's Functional Correctness}

In this section, we present SolBench, a benchmark for assessing the capabilities of Solidity code completion. Our approach emphasizes automatically checking the functional correctness of Solidity functions, which distinguishes it from earlier testing methods that relied on BLEU, CrystalBLEU scores~\cite{storhaug2023efficient}, or manual grading~\cite{dade2023optimizing} to evaluate the quality of generated smart contracts.

\subsection{Dataset Construction}
{The construction of SolBench consists of four steps}:
\paragraph{Step 1: Collection of Smart Contract Data.}
We adopt the DISL dataset~\cite{morello2024disl}, which is the largest dataset of Ethereum smart contracts. It contains a total of \num{514506} unique Solidity files that have been deployed on the Ethereum mainnet, representing real-world smart contract code. Since smart contracts on the blockchain are continuously being updated, our SolBench is also extensible.

\paragraph{Step 2: Extraction of Function-Level Code.}
We process the DISL dataset by extracting functions from the Solidity smart contracts it contains. Given the nature of our code completion task, we retain only functions that are accompanied by associated comments, defined structurally as comment blocks that immediately precede the function signature, following common Solidity documentation practices. Functions without such preceding comment blocks are excluded. These associated comments are then used to prompt our models to generate functions, which are subsequently evaluated for functional correctness.

\begin{figure}[t]
    \centering
    \includegraphics[width=0.9\columnwidth]{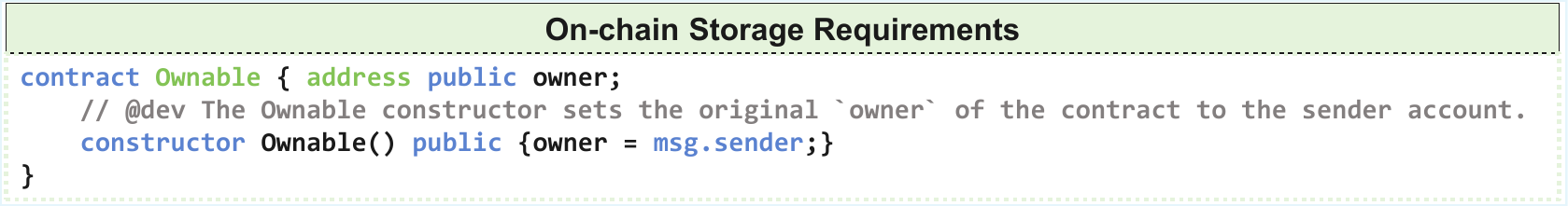} 
\caption{Smart contract example with on-chain storage for owner-only function execution (dataset construction Step~3). This figure illustrates why certain Oracle functions fail validation due to dependencies on on-chain storage, such as owner-only access control. These cases are filtered out during dataset construction to ensure accurate functional correctness verification.}

    \label{fig:on-chain-storage}
\end{figure}

\paragraph{Step 3: Verification of Oracle Solidity Functions for Functional Correctness.}
In this step, we need to verify that the Oracle Solidity functions we have extracted can pass functional correctness validation. This is because some functions call other functions within the same contract with the \texttt{only owner} modifier, which may require on-chain storage and cause the validation to fail. 
Unlike traditional stateless applications, which do not retain any information between executions, smart contracts need to maintain a record of their internal state to execute complex logic and maintain data integrity across transactions. This persistent state is stored on the blockchain, enabling the contract to track variables, balances, and other critical information over time. As shown in Fig.~\ref{fig:on-chain-storage}, the smart contract requires on-chain storage, the contract owner's address, to execute certain functions. Therefore, we further filtered functions that require on-chain storage for testing, such as those with constructors.
Additionally, some functions include mint operations, which are not allowed in our validation environment and would also lead to failure. Since we are using differential evaluation (see Section~\ref{sec-diffusc}), we add a statement \texttt{uint256 this\_is\_a\_test\_variable;} within the Oracle function body as a verification statement, as can be seen in Fig.~\ref{fig-echidna}. This creates a difference between the modified function and the original Oracle function, enabling the evaluator to identify the functions and associated variables that need to be verified. We retain the functions that pass the functional correctness validation.

\paragraph{Step 4: Deduplication of Function-Level Code and Dataset Construction.}
We deduplicate the verified functions based on entire functions, which include function comments, function signatures, and function bodies, to construct the SolBench. The rationale behind deduplicating at the function level is twofold: firstly, during model function completion, the prompt provided to the model consists of function comments and signatures, which could otherwise lead to the model generating identical responses. Secondly, this approach enhances the diversity of SolBench, ensuring a broader variety of examples.

\subsubsection{Implementation Details.} 
We started with the DISL dataset, which contains \num{514506} unique Solidity smart contracts. From this dataset, we extracted a total of \num{2609128} functions. However, after deduplicating these functions based on their full content, we obtained only \num{342975} unique functions, indicating that the original DISL dataset had a function duplication rate as high as 86.85\%.
Next, we validated each function to verify whether it could be compiled successfully and deployed correctly in our testing environment. We evaluated all \num{2609128} extracted functions for functional correctness, and \num{694953} of them passed this verification process.
Finally, we performed deduplication on the verified functions based on their complete content, resulting in a final set of \num{28825} functions across \num{7604} smart contracts, which constitutes our SolBench dataset.
Due to the high duplication rate of functions and the fact that many functions rely on on-chain storage, the resulting SolBench dataset is significantly smaller than the original DISL dataset.

\subsubsection{Dataset Statistics.}
\begin{table}[tbp]
  \centering
  
  \begin{minipage}{0.38\columnwidth}
    \centering
    \caption{
    Statistics of SolBench.
    }
    \small
    \setlength{\tabcolsep}{3pt}
    \begin{tabular}{lc}
        \toprule
        \textbf{Feature} & \textbf{Number} \\
        \midrule
        Func Count & 28825  \\
        SC Count & 7604  \\
        Uniq Func Sigs & 17936  \\
        Avg Func Lines & 15.25 \\
        Avg Func Body Lines & 9.4 \\
        Avg Func Body Tokens & 81.33 \\
        Avg Context Lines & 1126 \\
        Avg Context Tokens & 11654 \\
        \bottomrule
    \end{tabular}%
    \label{tab-SolBench-statistics}
    \end{minipage}%
\hspace{0.05\columnwidth}%
  \begin{minipage}{0.50\columnwidth}
    \centering
    \caption{Contract Type Statistics of SolBench}
    \scriptsize
    \setlength{\tabcolsep}{3pt}
    \resizebox{\linewidth}{!}{%
    \begin{tabular}{lccc}
        \toprule
        \textbf{Contract Type} 
        & \textbf{Func Count} 
        & \textbf{SC Count}
        & \textbf{Filtered (\%)} \\
        \midrule
        Token           & 12717 & 4323 & 89.00 \\
        NFT             & 6865  & 1409 & 88.50 \\
        DeFi            & 5486  & 1102 & 91.39 \\
        Governance      & 767   & 111  & 90.10 \\
        Application Logic& 736   & 138  & 90.57 \\
        Airdrop         & 579   & 147  & 91.86 \\
        Game            & 574   & 153  & 96.03 \\
        Smart Wallet    & 479   & 122  & 91.08 \\
        Marketplace     & 360   & 54   & 94.59 \\
        Smart Legal     & 262   & 45   & 94.12 \\
        \midrule
        Total Count                & 28825 & 7604 & -- \\
        \bottomrule
    \end{tabular}%
    }
    \label{tab-contract-type-statistics}
  \end{minipage}
  
\end{table}

As illustrated in Table~\ref{tab-SolBench-statistics}, we have systematically examined various data characteristics of the SolBench benchmark, providing a comprehensive overview. In SolBench, there are a total of \num{28825} functions for test distributed across \num{7604} contracts. After deduplication of all function signatures, there are \num{17936} unique function signatures, indicating that SolBench has a wide range of function coverage. Additionally, statistics were gathered on the average number of code lines for all functions, as well as the number of code lines and tokens required to complete the function bodies. We use the Qwen2.5-Coder Tokenizer to calculate the number of tokens. The average number of lines of context for samples in SolBench is \num{1126} lines, with a corresponding token count of \num{11654} tokens, demonstrating that the context is very long.
We further analyze potential data overlap by comparing SolBench against the Solidity subset of The Stack~\cite{Kocetkov2022TheStack}, the largest publicly available code corpus containing Solidity, and find that only 25.49\% of functions exhibit high similarity (Jaccard similarity $\geq 0.9$), suggesting limited overlap with existing public training data.

We also categorize the contracts in SolBench into ten common types based on major application scenarios of these contracts by providing their definitions to the GPT-4o-mini model, as shown in Table~\ref{tab-contract-type-statistics}. We further analyze the distribution of the functions and the contracts in SolBench across these categories. Among them, Token Contracts constitute the largest category, comprising \num{12717} functions across \num{4323} smart contracts, followed by NFT Contracts with \num{6865} functions and \num{1409} contracts. DeFi Contracts also represent a significant portion, containing \num{5486} functions from \num{1102} contracts. Beyond these dominant categories, the remaining contract types demonstrate the diversity of smart contract applications on Ethereum, covering a wide range of use cases such as governance mechanisms, gaming, airdrops, wallets, legal agreements, and marketplace functionalities. This variety highlights the diverse range of contract types included in SolBench, reflecting its comprehensive coverage of real-world Ethereum applications.

Finally, we investigate whether the filtering process disproportionately affects specific contract domains. To this end, we analyze the proportion of real-world, state-dependent functions removed by the filtering step across all contract categories in SolBench (Table~\ref{tab-contract-type-statistics}). We observe that the filtering process removes functions at a consistently high and comparable rate across different domains, generally around 90\%. In particular, Game, Marketplace, and Smart Legal contracts experience the highest removal rates, exceeding 94\%, while Token, NFT, and DeFi contracts exhibit similar filtering ratios of approximately 88--91\%. This relatively uniform filtering behavior across domains indicates that our filtering strategy does not introduce significant domain-specific coverage bias. Consequently, SolBench preserves the original diversity of smart contract types and remains representative of a broad spectrum of real-world Ethereum application scenarios.

\subsection{Verification of Functional Correctness}
\label{sec-diffusc}
In the following, we explain how to verify the functional correctness of the constructed SolBench dataset. We will first introduce the evaluation tools and methods, followed by a detailed workflow of these tools in the verification process.

\begin{figure}[tbp]
    \centering
    \includegraphics[width=0.8\textwidth]{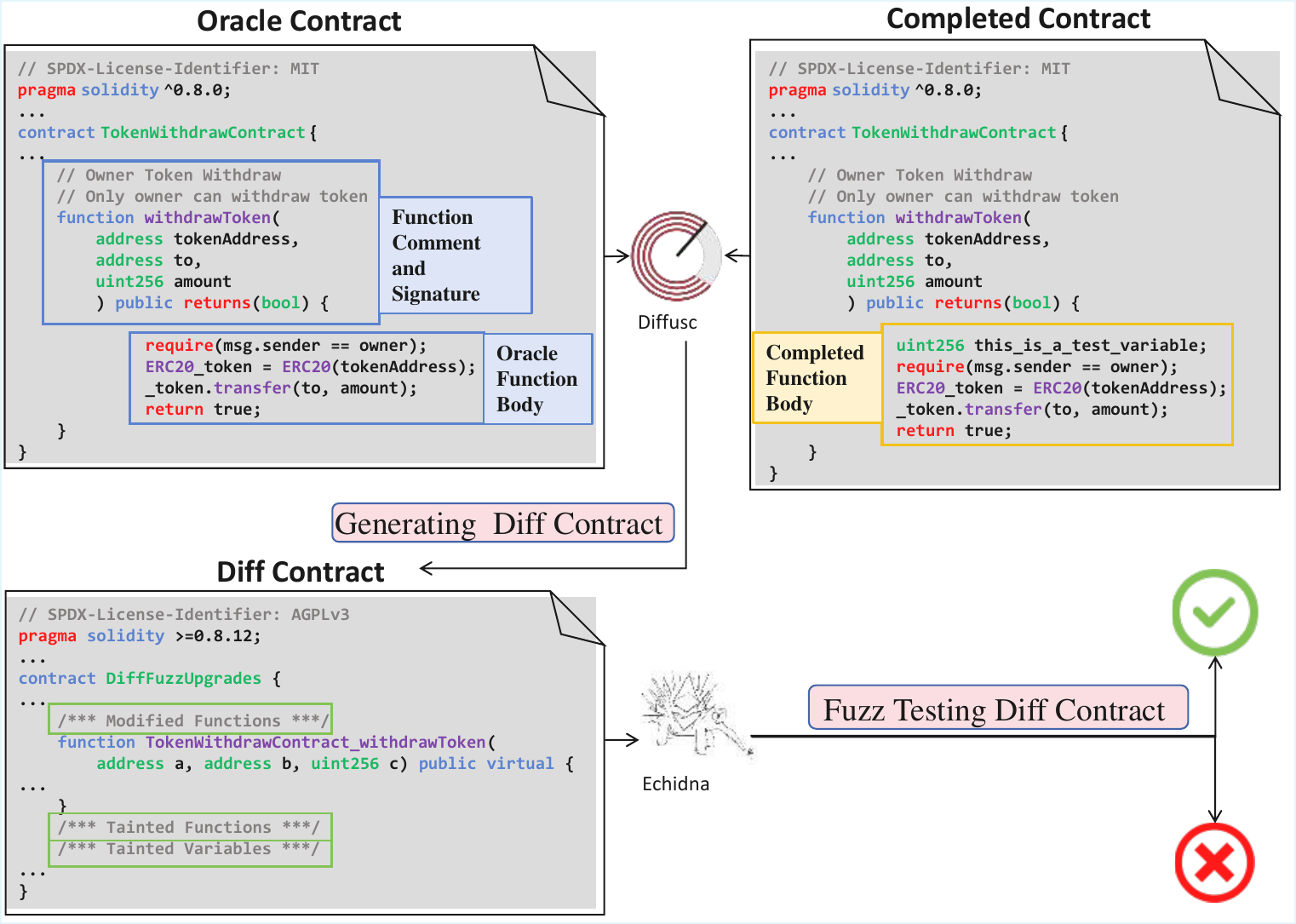} 

    \caption{
 Verification Process for SolBench Functional Correctness: Starting with an Oracle Contract from \texttt{SolBench}, the model completes the \texttt{withdrawToken} function based on the function comments and signature. Diffusc compares the Oracle and Completed Contracts, generating a Diff Contract. Echidna deploys the Diff Contract for fuzz testing on Modified Functions, Tainted Functions, and Variables.}
    \label{fig-echidna} 
\end{figure}

\subsubsection{Differential Fuzzing} 
Differential fuzzing is an advanced form of \textit{fuzz testing}, which is a security verification method that involves providing random inputs to software to identify vulnerabilities. In differential fuzzing, identical data is simultaneously input into two similar software implementations to identify any divergent behaviors in their execution. Throughout this process, the \textit{fuzzer}, a tool that injects random inputs, monitors the execution for any anomalies or errors.

\subsubsection{Diffusc}
Diffusc~\cite{CryticDiffusc2024} integrates static analysis with differential fuzz testing to scrutinize two versions of a smart contract prior to their on-chain deployment. Utilizing Slither~\cite{Feist_2019_Slither} frameworks, Diffusc employs differential taint analysis to create test contracts in Solidity tailored for differential fuzz testing. These contracts are subsequently subjected to fuzz testing via \textit{Echidna}~\cite{Grieco2020Echidna}. 

\subsubsection{Evaluation Protocols.} 
Solidity code completion aims to complete a function based on its comment and signature. After completing a function, we replace the original function in the smart contract. Following prior work~\cite{CryticDiffusc2024}, we use Diffusc to perform Differential Fuzzing to assess the functional correctness of the completed functions. To improve evaluation efficiency, we modify both Diffusc and Echidna to support multiprocessing, enabling concurrent verification of multiple functions. As shown in Fig.~\ref{fig-echidna}, for example:

\paragraph{Step 1.} {Starting with an Oracle Contract from \texttt{SolBench}, the model completes the \texttt{withdrawToken} function based on the function comments and signature.} This completed function is then substituted into the Oracle Contract's \texttt{withdrawToken} function to create a Completed Contract.
\paragraph{Step 2.} {Next, Diffusc compares the differences between the Oracle Contract and the Completed Contract, generating a Diff Contract.} Since the Oracle Function Body and the Completed Function Body in these two contracts differ, the Diff Contract generates a function called \texttt{TokenWithdrawContract\allowbreak\_withdrawToken} under the ``Modified Functions'' section. This function tests whether the \texttt{withdrawToken} functions in both contracts produce the same output given the same input. Additionally, the ``Tainted Functions'' and ``Tainted Variables'' sections generate functions to test any functions or variables directly invoked by \texttt{withdrawToken}, ensuring they also yield identical outputs for the same inputs.

\paragraph{Step 3.} Once the Diff Contract is generated, Echidna fuzz-tests all three sections: the modified functions, tainted functions, and tainted variables. If all functions pass the fuzz-testing, it confirms that the model's completion of the \texttt{withdrawToken} function is functionally correct.

    \subsection{Evaluation Metrics} 
\subsubsection{Pass Rate} 
Pass rate measures the percentage of functions completed by the models that not only compile and execute without errors but also produce the correct outputs for given inputs. This metric reflects the functional correctness of the completed functions, revealing the models' effectiveness in crafting smart contracts for practical use. We employ differential fuzzing, which means we treat the original function's output as the correct response. A completed function is considered successful if its output matches that of the original. We evaluate the pass rate of these functions in SolBench using the pass@k (P@k) metric, following the approach in~\cite{chen2021evaluating}:
\begin{equation}
    \text{pass}@k := \underset{\text{Problems}}{\mathbb{E}}\left[1 - \frac{\binom{n - c}{k}}{\binom{n}{k}}\right],
\end{equation}

where $n$ is the number of generated samples per problem, and $c$ is the number of correct solutions among them. In our evaluation, we compute pass@1 (P@1) by setting $n = 1$ and $k = 1$, which simplifies the formula to:
\begin{equation}
    \text{pass}@1 = \underset{\text{Problems}}{\mathbb{E}}\left[\mathbb{I}(c = 1)\right],
\end{equation}

i.e., the expected proportion of problems for which the single generated solution passes the functional correctness test. Results for larger values of pass@k are presented in Sec.~\ref{sec-larger-passk}.

\subsubsection{Compilation Rate} 
Compilation rate calculates the percentage of functions completed by the models that compile successfully without any errors, irrespective of their execution outcome. Although it's not the only factor, a high compilation rate is crucial for the code to be considered functionally correct. If a function compiles without issues, it's considered to have compiled correctly. We evaluate the compilation rate of these functions in SolBench using the compilation@1 (C@1) metric, similar to pass@1 metric.
\section{Benchmarking}

\subsection{Solidity Code Completion}

Using the constructed SolBench dataset and our evaluation pipeline for functional correctness, we assess LLMs’ Solidity code completion capabilities. We first consider a code completion setting in which, given only a function’s comments and signature within a contract, the model completes the function body. However, performance in this zero-context condition is poor: with a context length of 0 (Table~\ref{tab-llms-length}), GPT-4o-mini attains only 38.88\% P@1. Error analysis reveals that most failures stem from the model incorrectly using undeclared identifiers (Fig.~\ref{fig-error-distr}), underscoring that the task depends heavily on crucial details (e.g., type declarations and state variables) located elsewhere in the contract rather than in the function’s comments. These findings indicate that the task is highly context-dependent; accordingly, we evaluate LLM performance under different length of context.

We define the Solidity code completion task as: given a function’s comments and signature plus a certain length of preceding contract code as context, the model completes the function. Step~\textcircled{1} in Fig.~\ref{fig-rar-pipline} illustrates the Solidity code completion task. We measure context length in tokens using the Qwen2.5-Coder Tokenizer. We also try splitting the context evenly before and after the target function, but performance is slightly worse, likely because developers tend to define called functions earlier. Therefore, we use only the code that appears before the function as context.
\subsection{Experiments and Findings}
This section is guided by the following research questions:

\begin{itemize} 
\item \textbf{RQ1:} Why is functional correctness a necessary metric for smart contract code generation?
\item \textbf{RQ2:} How do LLMs perform on SolBench under varying context lengths?
\end{itemize}

\subsubsection{RQ1: Why is functional correctness a necessary metric for smart contract code generation?}
\label{sec-func-correct-matter}
To demonstrate that functional correctness (i.e., pass rate) is an essential metric for evaluating the correctness of Solidity code completion, we compare P@1 (functional correctness), BLEU, and CrystalBLEU scores for GPT-4o-mini across different context lengths. Table~\ref{tab-metric-compare} presents the results and the Pearson correlation coefficients between each metric and P@1. 
It is worth noting that all three metrics (P@1, BLEU, and CrystalBLEU) have scores from 0 to 100, which allows direct comparison and correlation analysis among them.

As shown in Table~\ref{tab-metric-compare}, P@1 increases rapidly with longer context, from 38.88 at context length 0 to 88.87 at 4k ($\Delta+49.99$). In contrast, BLEU and CrystalBLEU scores also improve with context length, but their growth is less pronounced: BLEU rises from 27.99 to 53.76 ($\Delta+25.77$), and CrystalBLEU rises from 19.11 to 47.35 ($\Delta+28.24$). More importantly, the correlation coefficients between BLEU/CrystalBLEU and P@1 are relatively low (BLEU: $r=0.26$–$0.41$; CrystalBLEU: $r=0.24$–$0.40$), indicating that these n-gram-based metrics do not strongly reflect the actual functional correctness of the generated code.

BLEU and CrystalBLEU are widely used in natural language generation tasks, but they primarily measure surface-level token overlap between the generated and reference code. In the context of code completion, especially for smart contracts, there can be multiple correct implementations that differ syntactically but are functionally equivalent. As a result, BLEU-based metrics may penalize valid solutions that use different variable names, code structures, or ordering, leading to a weak correlation with true correctness.
\begin{table}[tbp]
\centering
\caption{Comparison of P@1 (functional correctness), BLEU, and CrystalBLEU scores under different context lengths. “\textcolor{red}{Corr.}” indicates the Pearson correlation coefficient between BLEU/CrystalBLEU and P@1.}
\label{tab-metric-compare}
\small
\begin{tabular}{lcccccc}
\toprule
\textbf{\multirow{2}{*}{Metric}} & \multicolumn{6}{c}{\textbf{Context Length}} \\
\cmidrule(lr){2-7}
 & \textbf{0} & \textbf{256} & \textbf{512} & \textbf{1k} & \textbf{2k} & \textbf{4k} \\
\midrule
P@1 & 38.88 & 66.10 & 72.89 & 80.19 & 85.93 & 88.87 \\
BLEU / {\scriptsize\textcolor{red}{Corr.}} & 
27.99 / {\scriptsize\textcolor{red}{0.32}} & 
45.55 / {\scriptsize\textcolor{red}{0.41}} & 
48.39 / {\scriptsize\textcolor{red}{0.37}} & 
50.87 / {\scriptsize\textcolor{red}{0.34}} & 
52.92 / {\scriptsize\textcolor{red}{0.28}} & 
53.76 / {\scriptsize\textcolor{red}{0.26}} \\
CrystalBLEU / {\scriptsize\textcolor{red}{Corr.}} & 
19.11 / {\scriptsize\textcolor{red}{0.30}} & 
38.50 / {\scriptsize\textcolor{red}{0.40}} & 
41.59 / {\scriptsize\textcolor{red}{0.36}} & 
44.31 / {\scriptsize\textcolor{red}{0.32}} & 
46.45 / {\scriptsize\textcolor{red}{0.27}} & 
47.35 / {\scriptsize\textcolor{red}{0.24}} \\
\bottomrule
\end{tabular}
\end{table}

\begin{answer}
    \textbf{Answer to RQ1:} Functional correctness (P@1) is necessary because it directly captures whether generated Solidity code executes correctly, showing much greater sensitivity to context ($\Delta{+}49.99$ from 0 to 4k) than BLEU/CrystalBLEU. In contrast, BLEU-based metrics measure surface overlap and correlate weakly with correctness ($r{\approx}0.24$–$0.41$), missing functionally correct but syntactically different implementations.
\end{answer}

\subsubsection{RQ2: How do LLMs perform on SolBench under varying context lengths?}
In our experiments, we selected 14 large language models (Table~\ref{tab-selected-llms}) based on the following criteria: 
(a) varying accessibility, including open-source and closed-source models; 
(b) different scales, from lightweight (1.3B) to extremely large models (671B); 
(c) distinct training focuses, targeting either general-purpose or code-related tasks; 
(d) whether they are reasoning models
and (e) release dates covering both recent and earlier models. 
These criteria ensure a comprehensive and balanced evaluation. 
Models with $\leq 33$B parameters (DeepSeek-Coder, Qwen2.5-Coder, and CodeLlama) were deployed locally using vLLM on $4 \times$ NVIDIA A100 (80~GB) GPUs. All other models were accessed via APIs.
All models were sampled under their default inference settings. We do not test StarCoder2-15B-Instruct and WizardCoder-15B because these models were fine-tuned for Python instructions and performed poorly in our tests.
\begin{table}[tb]
\centering
\small
\caption{Summary of Selected LLMs.}
    \begin{tabular}{lcccc}
    \toprule
    \textbf{Model} & \textbf{Open-source} & \textbf{Code-specific} & \textbf{Reasoning model} & \textbf{Release time} \\
    \midrule
    Deepseek-Coder-1.3B-Instruct & $\checkmark$ & $\checkmark$ & $\times$ & Nov-23 \\ 
    Qwen2.5-Coder-1.5B-Instruct & $\checkmark$ & $\checkmark$ & $\times$ & Sep-24 \\ 
    CodeLlama-7B-Instruct & $\checkmark$ & $\checkmark$ & $\times$ & Aug-23 \\ 
    Deepseek-Coder-6.7B-Instruct & $\checkmark$ & $\checkmark$ & $\times$ & Nov-23 \\ 
    Qwen2.5-Coder-7B-Instruct & $\checkmark$ & $\checkmark$ & $\times$ & Sep-24 \\ 
    Deepseek-Coder-33B-Instruct & $\checkmark$ & $\checkmark$ & $\times$ & Nov-23 \\ 
    Qwen2.5-Coder-32B-Instruct & $\checkmark$ & $\checkmark$ & $\times$ & Sep-24 \\ 
    Deepseek-V3-671B (MoE) & $\checkmark$ & $\times$ & $\times$ & Dec-24 \\ 
    DeepSeek-R1-Distill-Llama-70B & $\checkmark$ & $\times$ & $\checkmark$ & Jan-25 \\ 
    Deepseek-R1-671B (MoE) & $\checkmark$ & $\times$ & $\checkmark$ & Jan-25 \\ 
    GPT-4o-mini & $\times$ & $\times$ & $\times$ & Oct-23 \\
    GPT-5-mini & $\times$ & $\times$ & $\checkmark$ & Aug-25 \\ 
    Doubao-Pro & $\times$ & $\times$ & $\times$ & Jan-25 \\ 
    Claude-3.5-Haiku & $\times$ & $\times$ & $\times$ & Oct-24 \\ 
    \bottomrule
    \end{tabular}
\label{tab-selected-llms}
\end{table}
\begin{table}[tbp]
\centering
\caption{The performance of different large models under various context lengths on SolBench. ``Avg'' measures a model's average performance across different context lengths. P@1 and C@1 denote pass@1 and compilation@1.}
\scriptsize 
\setlength{\tabcolsep}{3pt}
    \begin{tabular}{@{}lcccccccccccccc@{}}
        \toprule
        \textbf{\multirow{3}{*}{Model}} & \multicolumn{12}{c}{\textbf{Context Length}} & \multicolumn{2}{c}{} \\
        
        \cmidrule(lr){2-13}
        
         &     \multicolumn{2}{c}{\textbf{0}} &     \multicolumn{2}{c}{\textbf{256}} &     \multicolumn{2}{c}{\textbf{512}} &     \multicolumn{2}{c}{\textbf{1k}} &     \multicolumn{2}{c}{\textbf{2k}} &     \multicolumn{2}{c}{\textbf{4k}} & \multicolumn{2}{c}{\textbf{Avg.}} \\
         
         \cmidrule(lr){2-3} \cmidrule(lr){4-5} \cmidrule(lr){6-7}    \cmidrule(lr){8-9} \cmidrule(lr){10-11} \cmidrule(lr){12-13} \cmidrule(lr){14-15}
         
         & P@1 & C@1 & P@1 & C@1 & P@1 & C@1 & P@1 & C@1 & P@1 & C@1 & P@1 & C@1 & P@1 & C@1 \\
         
        \midrule
        
        \multicolumn{15}{c}{\textbf{Open-source Models}} \\
        Deepseek-Coder-1.3B-Instruct & 32.20 & 32.26 & 54.20 & 54.35 & 58.54 & 58.73 & 64.57 & 64.75 & 67.85 & 68.20 & 69.49 & 69.85 & 57.81 & 58.02 \\
        Qwen2.5-Coder-1.5B-Instruct & 39.53 & 39.60 & 55.16 & 55.33 & 60.59 & 60.77 & 66.10 & 66.33 & 70.09 & 70.41 & 71.77 & 72.13 & 60.54 & 60.76 \\
        CodeLlama-7B-Instruct & 21.73 & 21.73 & 45.71 & 45.84 & 52.81 & 52.94 & 59.53 & 59.71 & 57.23 & 57.45 & 57.37 & 57.58 & 49.06 & 49.21 \\
        Deepseek-Coder-6.7B-Instruct & \textbf{46.49} & 46.55 & 66.28 & 66.42 & 72.17 & 72.30 & 77.63 & 77.79 & 83.04 & 83.31 & 85.36 & 85.67 & 71.83 & 72.01 \\
        Qwen2.5-Coder-7B-Instruct & 43.51 & 43.53 & 65.30 & 65.38 & 70.24 & 70.33 & 75.88 & 75.99 & 81.03 & 81.26 & 82.88 & 83.10 & 69.81 & 69.93 \\
        Deepseek-Coder-33B-Instruct & 46.18 & 46.23 & 67.98 & 68.10 & 72.97 & 73.10 & 78.74 & 78.91 & 84.04 & 84.28 & 86.60 & 86.86 & 72.75 & 72.91 \\
        Qwen2.5-Coder-32B-Instruct & 42.18 & 42.21 & 64.76 & 64.84 & 71.65 & 71.75 & 77.46 & 77.58 & 83.29 & 83.49 & 86.27 & 86.51 & 70.94 & 71.06 \\
        Deepseek-V3-671B (MoE) & 44.71 & 44.73 & \textbf{71.21} & 71.28 & \textbf{77.57} & 77.66 & \textbf{84.44} & 84.54 & \textbf{90.05} & 90.24 & \textbf{92.82} & 93.05 & \textbf{76.80} & 76.92 \\
        DeepSeek-R1-Distill-Llama-70B & 23.06 & 24.85 & 47.10 & 50.93 & 53.23 & 57.62 & 59.91 & 65.05 & 63.94 & 69.62 & 66.35 & 72.18 & 52.27 & 56.71 \\
        Deepseek-R1-671B (MoE) & 42.90 & 45.48 & 67.95 & 72.01 & 74.71 & 78.86 & 81.15 & 85.82 & 86.14 & 91.03 & 88.28 & 93.41 & 73.52 & 77.77 \\

        \midrule
        \multicolumn{15}{c}{\textbf{Closed-source Models}} \\
        GPT-4o-mini & 38.88 & 38.95 & 66.10 & 66.20 & 72.89 & 73.01 & 80.19 & 80.33 & 85.93 & 86.21 & 88.87 & 89.20 & 72.14 & 72.32 \\
        GPT-5-mini & \textbf{51.90} & 51.93 & \textbf{77.31} & 77.47 & \textbf{83.01} & 83.20 & \textbf{89.17} & 89.38 & \textbf{93.61} & 93.96 & \textbf{95.68} & 96.02 & 81.78 & 81.99 \\
        Doubao-1.5-pro & 39.21 & 39.24 & 66.16 & 66.24 & 72.25 & 72.37 & 80.22 & 80.33 & 85.78 & 86.03 & 88.75 & 88.97 & 72.06 & 72.20 \\
        Claude-3.5-Haiku & 47.98 & 48.03 & 71.59 & 71.68 & 77.96 & 78.10 & 84.95 & 85.10 & 90.67 & 90.90 & 93.47 & 93.71 & 77.77 & 77.92 \\
        
        \bottomrule
    \end{tabular}
\label{tab-llms-length}
\end{table}

As shown in Table~\ref{tab-llms-length}, increasing context length consistently improves P@1 and C@1 across all models, with peak performance at 4k context. At 4k, GPT-4o-mini attains an 88.87\% P@1, compared to just 38.88\% with a context length of 0, underscoring that access to contextual information is the primary bottleneck for Solidity code completion.
Among closed-source models, GPT-5-mini achieves the highest performance with an average P@1 of 81.78\% and C@1 of 81.99\%, outperforming the strongest open-source model Deepseek-V3-671B (P@1: 76.80\%, C@1: 76.92\%).
Within model series like Deepseek-Coder and Qwen2.5-Coder, larger sizes yield better results. Smaller models (such as Deepseek-Coder-1.3B-Instruct and Qwen2.5-Coder-1.5B-Instruct) struggle with long-context completion tasks. However, the 7B models already achieve strong performance (e.g., Deepseek-Coder-6.7B-Instruct: 71.83\% P@1, 72.01\% C@1), with only marginal gains from scaling up to 33B or 32B. Deepseek-Coder-33B-Instruct is the best-performing code-specific model (P@1: 72.75\%, C@1: 72.91\%), comparable to some closed models, highlighting the effectiveness of code-focused fine-tuning.
Reasoning-oriented models like DeepSeek-R1 show no advantage and even slight drops compared to Deepseek-V3, likely because SolBench’s tasks are relatively simple compared with competition-level code or mathematical problems, where excessive reasoning may hurt performance.
Finally, newer models clearly outperform older ones. Despite smaller size, Deepseek-Coder-1.3B and Qwen2.5-Coder-1.5B surpass CodeLlama-7B in both P@1 and C@1, showing the benefit of more recent training recipes.
\begin{answer}
    \textbf{Finding 1:} Performance increases with context length, and larger, newer or code-tuned models perform better. GPT-5-mini achieves the best overall results; among open-source models, DeepSeek-V3 leads. Small models struggle on long contexts and reasoning-oriented variants offer no gains.
\end{answer}

\paragraph{Model Performance under Longer Context Lengths}
\begin{table}[tbp]
\centering
\caption{The performance of LLMs under longer context lengths on SolBench. P@1 and C@1 denote pass@1 and compilation@1.}
\small
    \begin{tabular}{@{}lcccccc@{}}
        \toprule
        \textbf{\multirow{3}{*}{Model}} & \multicolumn{6}{c}{\textbf{Context Length}} \\
        
        \cmidrule(lr){2-7}
        
         &     \multicolumn{2}{c}{\textbf{8k}} &     \multicolumn{2}{c}{\textbf{16k}} &     \multicolumn{2}{c}{\textbf{32k}} \\
         
         \cmidrule(lr){2-3} \cmidrule(lr){4-5} \cmidrule(lr){6-7}
         
         & P@1 & C@1 & P@1 & C@1 & P@1 & C@1 \\
         
        \midrule
        
        Deepseek-V3-671B (MoE) & 93.94 & 94.16 & 94.13 & 94.36 & 94.14 & 94.37 \\

        Deepseek-R1-671B (MoE) & 92.42 & 92.66 & 92.63 & 92.88 & 92.60 & 92.84 \\

        GPT-4o-mini & 90.76 & 91.06 & 91.34 & 91.66 & 91.67 & 91.98 \\

        GPT-5-mini & 96.69 & 97.03 & 96.99 & 97.37 & 97.19 & 97.56 \\

        Doubao-1.5-pro & 89.58 & 89.86 & 89.99 & 90.30 & 89.24 & 89.55 \\
        
        Claude-3.5-Haiku & 94.57 & 94.84 & 95.03 & 95.29 & 95.12 & 95.37 \\

        \bottomrule
    \end{tabular}
\label{tab-llms-longer-length}
\end{table}
To further investigate the Solidity code completion capabilities of LLMs under extended context lengths, we selected several models that support longer context windows and evaluated their performance at context lengths ranging from 8k to 32k. We excluded the Deepseek-Coder and Qwen2.5-Coder series from these experiments because their maximum supported context lengths are 16k and 32k, respectively, and some cases exceeded these limits. Additionally, DeepSeek-R1-Distill-Llama-70B was not included due to computational resource constraints. Notably, 99.16\% of functions in the SolBench dataset have context lengths within 32k, so the results at this length effectively reflect model performance when provided with the complete available context.

As shown in Table~\ref{tab-llms-longer-length}, GPT-5-mini achieves a P@1 score of 97.19\% at 32k, only a modest improvement over 95.68\% at 4k. This supports the consensus that expanding the context window beyond a certain point yields diminishing returns, as models struggle to fully utilize very long contexts.
\begin{answer}
    \textbf{Finding 2:} Although longer contexts provide some benefit, the gains diminish as context length increases.
\end{answer}

\paragraph{Model Performance under Different Pass@k}
\begin{table}[tbp]
\centering
\caption{The performance of LLMs under different pass@k on SolBench.}
\small
    \begin{tabular}{@{}lcccccccc@{}}
        \toprule
        \textbf{\multirow{3}{*}{Model}} & \multicolumn{8}{c}{\textbf{Context Length}} \\
        
        \cmidrule(lr){2-9}
        
         &     \multicolumn{4}{c}{\textbf{256}} &     \multicolumn{4}{c}{\textbf{1k}}\\
         
         \cmidrule(lr){2-5} \cmidrule(lr){6-9}
         
         & P@1 & P@2 & P@3 & P@5 & P@1 & P@2 & P@3 & P@5 \\
         
        \midrule
        
        Deepseek-V3-671B (MoE) & 71.45 & 73.51 & 74.52 & 75.66 & 84.58 & 86.34 & 87.12 & 87.97 \\

        GPT-4o-mini & 64.84 & 69.53 & 71.70 & 74.03 & 79.08 & 83.09 & 84.77 & 86.53 \\
        \bottomrule
    \end{tabular}
\label{tab-llms-length-passk}
\end{table}
\label{sec-larger-passk}
To further evaluate the code completion capabilities of large language models, we selected $n=5$ and tested model performance under different $k$ values (i.e., Pass@$k$).

As shown in Table~\ref{tab-llms-length-passk}, under a context length of 256, GPT-4o-mini achieves a P@1 of 64.84\%, which increases to 74.03\% at P@5. Similarly, Deepseek-V3-671B achieves a P@1 of 71.45\%, rising to 75.66\% at P@5. However, this trend is less pronounced with longer context lengths: for a context length of 1k, GPT-4o-mini improves from 79.08\% (P@1) to 86.53\% (P@5), and Deepseek-V3-671B from 84.58\% (P@1) to 87.97\% (P@5). These results indicate that while allowing the model to generate more candidate completions can effectively increase its pass rate in code completion tasks, the marginal benefit diminishes as the context length increases. This is because longer contexts contain more information relevant to the completion, making the target function more deterministic.
\begin{answer}
    \textbf{Finding 3:} Allowing higher k substantially increases pass rates at short context lengths, but the incremental gains shrink as context grows (e.g., 256→1k). 
\end{answer}

\paragraph{Error Distributions.} 
\begin{figure}[tbp]
\centering
\includegraphics[width=0.48\textwidth]{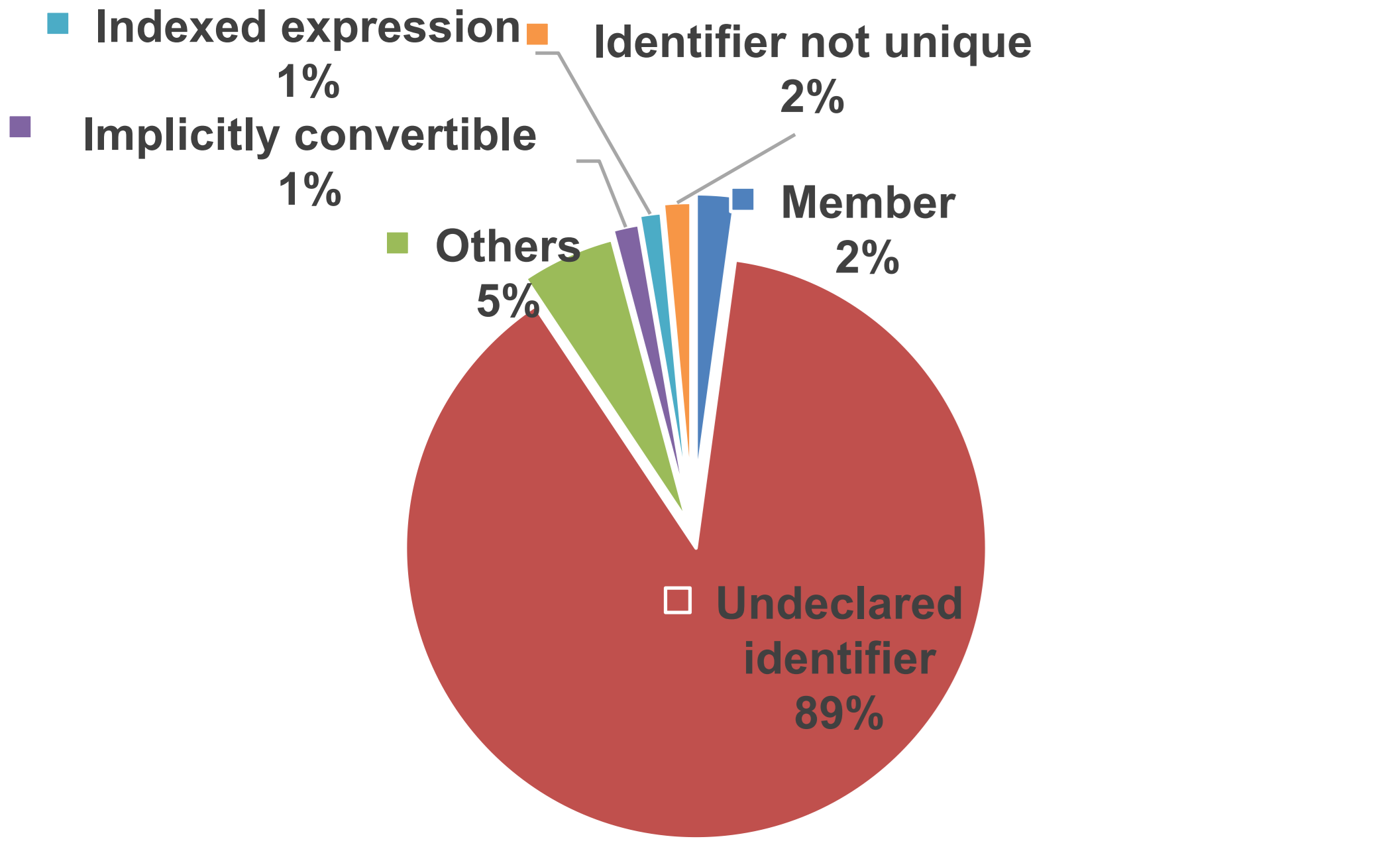} 
\caption{
Top five error distributions for code completion when the context length is 0 using GPT-4o on SolEval. The majority of the errors are ``Undeclared Identifier'', meaning the model's function completions include identifiers that are not declared.
} 
\label{fig-error-distr} 
\end{figure}

We have analyzed the top five error distributions for GPT-4o-mini when the context length is 0 on SolBench (Fig.~\ref{fig-error-distr}). The majority of the errors are ``Undeclared Identifier'', meaning the model's function completions include identifiers that are not declared. The remaining errors account for a much smaller proportion: ``Member'' errors occur when the model generates struct types or other structures with non-existent member names; ``Identifier not unique'' errors arise when there is ambiguity in the definitions of generated identifiers; ``Indexed expression'' errors happen when the model incorrectly indexes objects that cannot be indexed; and ``Implicitly convertible'' errors occur when the model uses identifiers of incorrect types.

\subsection{Analysis}
Solidity code completion is difficult for LLMs because it depends on crucial details, such as type definitions and state variables, that usually reside elsewhere in the contract rather than in the incomplete function’s comments.
Without these critical details, models often produce plausible yet incorrect code due to insufficient context.
For example, models might use incorrect variables or call wrong functions, leading to compilation errors (Fig.~\ref{fig-error-distr}). 
Including the entire contract as model input can provide the necessary context, but this approach dramatically increases computational cost. 
Transformer-based LLMs (e.g., GPT) exhibit quadratic growth in memory and computation as context length increases, making long-context inference slow and expensive.
This challenge is pronounced in SolBench, where the average full contract spans \num{11654} tokens per sample (Table~\ref{tab-SolBench-statistics}).
At current API pricing, a full-context SolBench run with Claude-Opus-4.1 would cost about \$5,286, illustrating that directly scaling context is economically impractical.
Yet most of the contract isn’t needed to implement the current function; we only need the crucial details relevant to that function. Therefore, a precise extraction of only the useful code snippets can complete the function while greatly reducing inference cost.
\section{Retrieval-Augmented Repair}
\begin{figure*}[t]
    \centering
    \includegraphics[width=0.9\textwidth]{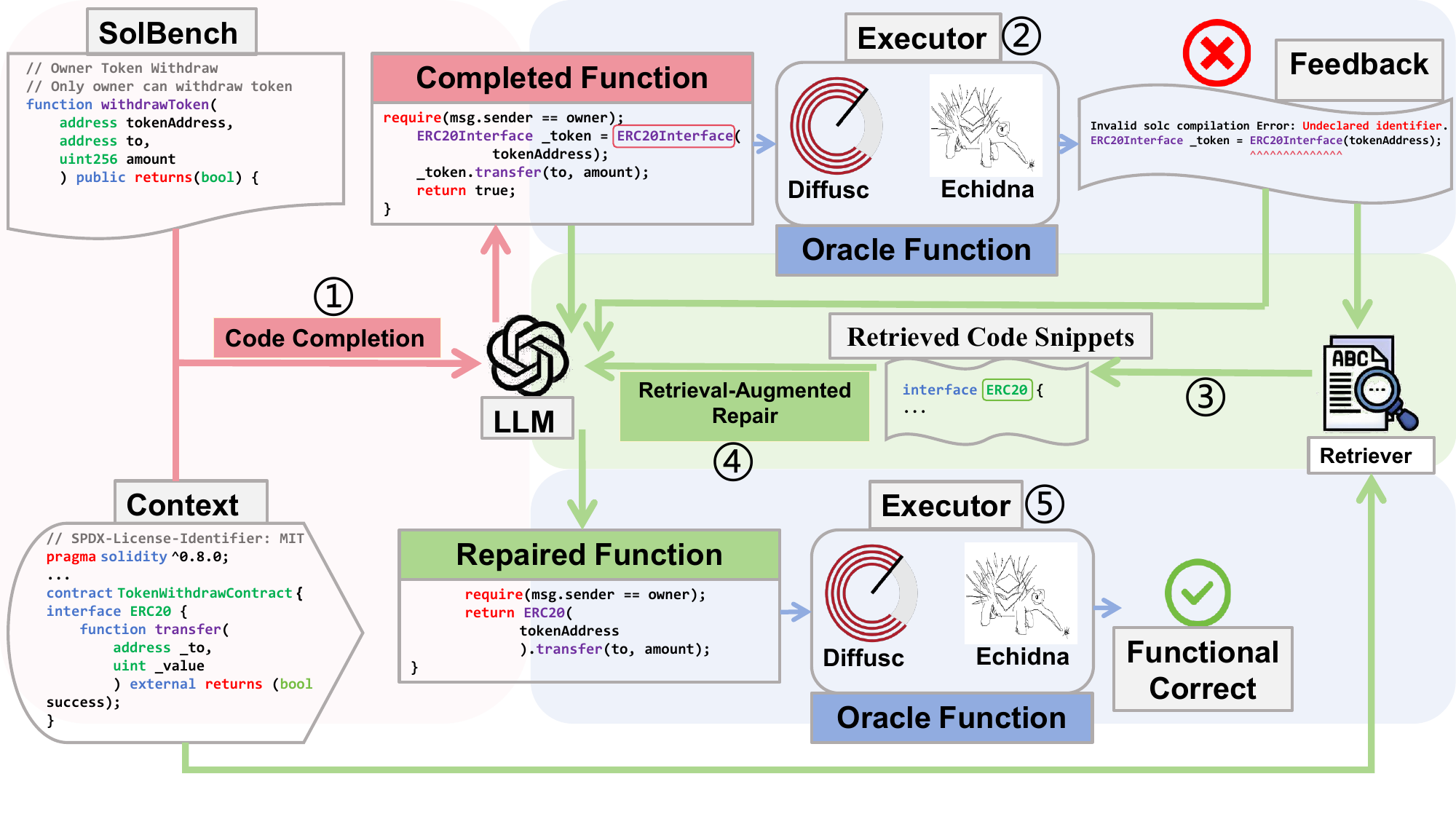}

    \caption{Illustration of our proposed
Retrieval-Augmented Repair (RAR) framework: \textcircled{1} The LLM first completes a SolBench function using the provided context (i.e., code completion). \textcircled{2} The executor then verifies the functional correctness of the completed function by comparing it to the oracle function. \textcircled{3} If the function is incorrect, a retriever retrieves relevant code snippets from the context based on feedback from the executor. \textcircled{4} The LLM then repairs the incorrect function by leveraging both the executor’s feedback and the retrieved code snippets (i.e., retrieval-augmented repair). \textcircled{5} Finally, the executor verifies the functional correctness of the repaired function.
}
    \label{fig-rar-pipline} 
\end{figure*}

\subsection{Our Proposed Retrieval-Augmented Repair Framework}
\label{sec:rag}
To address the dual challenges of (i) insufficient context in incomplete functions, which leads models to misuse types, variables, or function calls, and (ii) the prohibitive compute and monetary cost of supplying full contract context due to quadratic scaling with long sequences (especially acute in the long contracts of SolBench), we introduce the Retrieval-Augmented Repair (RAR) Framework (Fig.~\ref{fig-rar-pipline}).
RAR integrates retrieval into code repair, using the executor’s error messages to extract only the most relevant crucial details from the full contract. This approach sharply reduces input length for function completion, improving code completion accuracy while significantly cutting computational cost.
The workflow of Retrieval-Augmented Repair (RAR) is shown in Fig.~\ref{fig-rar-pipline}. The LLM first completes a Solidity function. If the executor detects an error, it returns a message identifying the faulty line. This line is then used as a query to retrieve relevant code snippets from the broader function context, producing the \emph{Retrieved Code Snippets}. Finally, the LLM repairs the function by leveraging both the executor’s feedback and the retrieved snippets.
\subsection{Code Repair and Retrieval Methods}
        \subsubsection{Code Repair Methods}
Code repair is a prompting strategy widely adopted in prior work. In our evaluation, we consider several commonly used code repair methods as baselines for the proposed RAR framework. Following established practices in LLM-based code repair, we select four representative strategies: three execution-based methods (Self Edit, Self Debug, and Self Repair) and one reasoning-based method (Self Refine). Together, these baselines cover the major prompting paradigms explored in existing literature.

\begin{itemize}[leftmargin=*, itemsep=0pt, topsep=0pt]
    \item Self Edit~\cite{zhang2023selfeditfaultawarecodeeditor} directly uses the error information returned by the executor as context to prompt the model for code repair.
    \item Self Debug~\cite{chen2023teachinglargelanguagemodels} first asks the model to explain each line of the completed function, then uses the error information returned by the executor as context to prompt the model for code repair.
    \item Self Refine~\cite{madaan2023selfrefineiterativerefinementselffeedback} does not use the error information returned by the executor but instead asks the model to reflect on where the completed function went wrong, and then performs code repair based on the inferred error.
    \item Self Repair~\cite{olausson2024selfrepairsilverbulletcode} first asks the model to interpret the meaning of the error information returned by the executor and then performs code repair.
\end{itemize}

    \subsubsection{Retrieval Methods}
RAR further enhances code repair by retrieving relevant information from the broader context of an incomplete Solidity function. In our RAR framework, we evaluate both sparse and dense retrieval strategies following established practices in retrieval-augmented code generation. Specifically, we select six widely used retrieval methods as baselines, including four sparse approaches (LCS, BM25, TF-IDF, and Jaccard Similarity) and two dense approaches (UniXCoder and CodeBERT).
\paragraph{Sparse Retrieval Methods}
Sparse retrieval techniques include:
\begin{itemize}[leftmargin=*, itemsep=0pt, topsep=0pt]
    \item LCS (Longest Common Substring)~\cite{li2024codesbuildingopensourcelanguage}: This algorithm identifies the longest common substring between two given strings. In our approach, we use the undeclared identifier from the error message returned by the executor as the query, and start matching from the longest substring of the query. If a substring matches a certain code line, we return that code line. If the error message does not specify an identifier, we parse all identifiers in the error message or the initially completed function using a Solidity parser as the query. We test the LCS algorithm because the main error in Solidity code completion is the \emph{Undeclared Identifier Error} (Fig.~\ref{fig-error-distr}), where the model generates similar but incorrect identifiers. As shown in Fig.~\ref{fig-rar-pipline}, the LCS algorithm effectively matches the model-generated identifiers to the correct ones in the context. 

    \item BM25~\cite{ding2024crosscodeeval}: A probabilistic ranking function widely used in information retrieval tasks to rank documents based on their relevance to a query. In our experiment, BM25 is used to evaluate the relevance of code lines to the query derived from the error message.
    
    \item TF-IDF (Term Frequency-Inverse Document Frequency): A statistical method that assesses the importance of a term within a document or corpus. Here, TF-IDF is employed to weigh the significance of terms in the error message.
    
    \item Jaccard Similarity~\cite{zhang2023repocoder}: A measure used to calculate the similarity between two sets. In this scenario, Jaccard Similarity evaluates the overlap between the set of terms in the query (e.g., identifiers from the error message) and the set of terms in each code line. Code lines with higher similarity scores are considered more relevant to the query.
\end{itemize}
    
\paragraph{Dense Retrieval Methods}
Dense retrieval harnesses the power of encoder models to produce code embeddings that capture the underlying meaning of code snippets. We use \emph{cosine similarity} to assess the semantic similarity between a query and candidate code snippets. We select these encoder models:

\begin{itemize}[leftmargin=*, itemsep=0pt, topsep=0pt]
    \item UniXCoder~\cite{guo2022unixcoder}: We utilize UnixCoder-base, an encoder-only unified cross-modal pre-trained model that leverages multimodal data (i.e., code comments and AST) to pretrain code representation.
    \item CodeBERT~\cite{feng2020codebert}: We use the CodeBERT-base version, a bimodal pre-trained model for programming language and natural language with a total of 125M parameters.
\end{itemize}

For all retrieval methods other than LCS, we use the erroneous code line from the executor's error message as the query to find the most relevant code snippet in the context.

\subsection{Experiments and Findings}
This section addresses the following research questions:
\begin{itemize} 
\item \textbf{RQ3:} How does the RAR framework perform with different retrievers and context lengths?
\item \textbf{RQ4:} How effective is RAR in reducing inference cost?
\end{itemize}

\subsubsection{RQ3: How does the RAR framework perform with different retrievers and context lengths?}
        \paragraph{Assessing Performance of Baseline Code Repair Methods}
\begin{table}[tb]
    \caption{Evaluation of Code Repair Methods on SolBench Using GPT-4o-mini. ``No Repair'' refers to single generation without repair. ``Avg.'' shows the average performance across different context lengths. ``Avg.\ CR'' measures average performance for repair methods at specific lengths, and ``Increase'' indicates improvement over ``No Repair''. P@1 and C@1 denote pass@1 and compilation@1.}
    \centering
    \scriptsize 
    \begin{tabular}{@{}lcccccccccccccc@{}}
        \toprule
        \textbf{\multirow{3}{*}{CR Method}} & \multicolumn{12}{c}{\textbf{Context Length}} & \multicolumn{2}{c}{} \\
        \cmidrule(lr){2-13}
         &     \multicolumn{2}{c}{\textbf{0}} &     \multicolumn{2}{c}{\textbf{256}} &     \multicolumn{2}{c}{\textbf{512}} &     \multicolumn{2}{c}{\textbf{1k}} &     \multicolumn{2}{c}{\textbf{2k}} &     \multicolumn{2}{c}{\textbf{4k}} & \multicolumn{2}{c}{\textbf{Avg.}} \\
         \cmidrule(lr){2-3} \cmidrule(lr){4-5} \cmidrule(lr){6-7}    \cmidrule(lr){8-9} \cmidrule(lr){10-11} \cmidrule(lr){12-13} \cmidrule(lr){14-15}
         & P@1 & C@1 & P@1 & C@1 & P@1 & C@1 & P@1 & C@1 & P@1 & C@1 & P@1 & C@1 & P@1 & C@1 \\
        \midrule
        No Repair      & 38.88 & 38.95 & 66.10 & 66.20 & 72.89 & 73.01 & 80.19 & 80.33 & 85.93 & 86.21 & 88.87 & 89.20 & 72.14 & 72.32 \\
        Self Edit      & 45.85 & 45.97 & 70.55 & 70.73 & 76.72 & 76.93 & 83.43 & 83.67 & 88.71 & 89.20 & 91.39 & 91.95 & 76.11 & 76.41 \\
        Self Debug & \textbf{47.38} & 47.51 & \textbf{71.77} & 71.94 & \textbf{77.87} & 78.07 & \textbf{83.90} & 84.12 & \textbf{89.29} & 89.75 & \textbf{91.47} & 91.98 & \textbf{76.95} & 77.23 \\
        Self Repair    & 46.75 & 46.87 & 70.85 & 71.01 & 76.63 & 76.81 & 83.00 & 83.21 & 88.31 & 88.79 & 91.01 & 91.58 & 76.09 & 76.38  \\
        Self Refine    & 39.72 & 39.85 & 66.94 & 67.11 & 73.62 & 73.80 & 80.82 & 81.05 & 86.62 & 87.11 & 89.69 & 90.23 & 72.90 & 73.19  \\
        \hline
        {Avg. CR}      & 44.93 & 45.05 & 70.03 & 70.20 & 76.21 & 76.40 & 82.79 & 83.01 & 88.23 & 88.71 & 90.89 & 91.44 & 75.51 & 75.80  \\
        Increase (\%)  & 6.05 & 6.10 & 3.93 & 4.00 & 3.32 & 3.39 & 2.60 & 2.68 & 2.30 & 2.50 & 2.02 & 2.24 & 3.37 & 3.49   \\
        \bottomrule        
    \end{tabular}
    \label{tab-cot-length}
\end{table}

We tested the extent to which different code repair methods improve the pass rate under varying context lengths. We used GPT-4o-mini as both the code completion and repair model due to its favorable cost--performance trade-off. While GPT-5-mini and Claude-3.5-Haiku achieve the top performance among the evaluated models (Table~\ref{tab-llms-length}), under our inference setup (Section~\ref{sec-cost-effective}), GPT-5-mini incurs approximately 88\% higher average cost than GPT-4o-mini, and Claude-3.5-Haiku is 465\% more expensive. Therefore, GPT-4o-mini offers a practical balance between performance and computational cost for large-scale RAR experiments.

As shown in Table~\ref{tab-cot-length}, all repair methods yield higher performance compared to direct generation without repair, and the improvements are more significant when the available context is limited. At context length 0, applying code repair increases the average P@1 and C@1 by 6.05\% and 6.10\%, respectively. This highlights the effectiveness of code repair in low-context settings where generation is most challenging.
Among all methods, \emph{Self Debug} achieves the highest overall performance, with the average P@1 of 76.95\% and C@1 of 77.23\%. Similarly, \emph{Self Edit} and \emph{Self Repair}—both incorporating executor feedback—show strong and stable improvements, closely trailing \emph{Self Debug} in both metrics. In contrast, \emph{Self Refine}, which relies solely on the model’s self-reflection without external feedback, exhibits the weakest performance, with average pass and compilation rates of 72.90\% and 73.19\%, respectively. This performance gap emphasizes that external error feedback is more reliable than internal reflection for effective code correction.
\begin{answer}
    \textbf{Finding 1:} Under varying context lengths, code repair methods consistently improve both pass rate and compilation rate, with more pronounced gains under shorter contexts.
\end{answer}

        \paragraph{Assessing Performance Under Different Retrieval Methods}
        \begin{table}[tb]
\caption{Pass@1 performance of Retrieval-Augmented Repair on SolBench using GPT-4o-mini with context lengths of 256 and 1k. ``No Retrieval'' denotes the baseline code repair without external retrieval. ``Average'' indicates the average pass rate across repair methods. $\uparrow x.xx$ indicates the absolute Pass@1 improvement over the ``No Retrieval'' baseline.}

\centering
\scriptsize
\begin{tabular}{>{\bfseries}lccccccc}
\toprule
\multirow{2}{*}{\textbf{CR Method}} & \multicolumn{5}{c}{\textbf{Retrieval Method}} & \multicolumn{2}{c}{\textbf{Context Length = 256}} \\
\cmidrule(l){2-8}
& \textbf{No Retrieval} & \textbf{LCS} & \textbf{BM25} & \textbf{TF-IDF} & \textbf{Jaccard-Sim} & \textbf{UniXCoder} & \textbf{CodeBERT} \\
\midrule
Self Edit
& 70.63
& 77.59 {\scriptsize\textcolor{gray}{$\uparrow$6.96}}
& 77.88 {\scriptsize\textcolor{gray}{$\uparrow$7.25}}
& 77.70 {\scriptsize\textcolor{gray}{$\uparrow$7.07}}
& 77.74 {\scriptsize\textcolor{gray}{$\uparrow$7.11}}
& 78.29 {\scriptsize\textcolor{gray}{$\uparrow$7.66}}
& 77.59 {\scriptsize\textcolor{gray}{$\uparrow$6.96}} \\

Self Debug
& 71.84
& 75.13 {\scriptsize\textcolor{gray}{$\uparrow$3.29}}
& 73.91 {\scriptsize\textcolor{gray}{$\uparrow$2.07}}
& 73.84 {\scriptsize\textcolor{gray}{$\uparrow$2.00}}
& 73.75 {\scriptsize\textcolor{gray}{$\uparrow$1.91}}
& 74.75 {\scriptsize\textcolor{gray}{$\uparrow$2.91}}
& 74.01 {\scriptsize\textcolor{gray}{$\uparrow$2.17}} \\

Self Repair
& 70.91
& 74.45 {\scriptsize\textcolor{gray}{$\uparrow$3.54}}
& 73.20 {\scriptsize\textcolor{gray}{$\uparrow$2.29}}
& 73.15 {\scriptsize\textcolor{gray}{$\uparrow$2.24}}
& 73.08 {\scriptsize\textcolor{gray}{$\uparrow$2.17}}
& 73.47 {\scriptsize\textcolor{gray}{$\uparrow$2.56}}
& 72.93 {\scriptsize\textcolor{gray}{$\uparrow$2.02}} \\

Self Refine
& 67.01
& 72.01 {\scriptsize\textcolor{gray}{$\uparrow$5.00}}
& 69.82 {\scriptsize\textcolor{gray}{$\uparrow$2.81}}
& 69.85 {\scriptsize\textcolor{gray}{$\uparrow$2.84}}
& 69.35 {\scriptsize\textcolor{gray}{$\uparrow$2.34}}
& 70.47 {\scriptsize\textcolor{gray}{$\uparrow$3.46}}
& 69.54 {\scriptsize\textcolor{gray}{$\uparrow$2.53}} \\

\midrule
Average
& 70.10
& \textbf{74.80} {\scriptsize\textcolor{gray}{$\uparrow$\textbf{4.70}}}
& 73.70 {\scriptsize\textcolor{gray}{$\uparrow$3.60}}
& 73.64 {\scriptsize\textcolor{gray}{$\uparrow$3.54}}
& 73.48 {\scriptsize\textcolor{gray}{$\uparrow$3.38}}
& 74.25 {\scriptsize\textcolor{gray}{$\uparrow$4.15}}
& 73.52 {\scriptsize\textcolor{gray}{$\uparrow$3.42}} \\

\toprule
{\textbf{}} & \multicolumn{5}{c}{\textbf{}} & \multicolumn{2}{c}{\textbf{Context Length = 1k}} \\
\midrule
Self Edit
& 83.53
& 86.99 {\scriptsize\textcolor{gray}{$\uparrow$3.46}}
& 87.09 {\scriptsize\textcolor{gray}{$\uparrow$3.56}}
& 86.99 {\scriptsize\textcolor{gray}{$\uparrow$3.46}}
& 87.05 {\scriptsize\textcolor{gray}{$\uparrow$3.52}}
& 87.15 {\scriptsize\textcolor{gray}{$\uparrow$3.62}}
& 86.97 {\scriptsize\textcolor{gray}{$\uparrow$3.44}} \\

Self Debug
& 83.98
& 85.60 {\scriptsize\textcolor{gray}{$\uparrow$1.62}}
& 84.82 {\scriptsize\textcolor{gray}{$\uparrow$0.84}}
& 84.69 {\scriptsize\textcolor{gray}{$\uparrow$0.71}}
& 84.87 {\scriptsize\textcolor{gray}{$\uparrow$0.89}}
& 85.01 {\scriptsize\textcolor{gray}{$\uparrow$1.03}}
& 84.81 {\scriptsize\textcolor{gray}{$\uparrow$0.83}} \\

Self Repair
& 83.07
& 85.25 {\scriptsize\textcolor{gray}{$\uparrow$2.18}}
& 84.62 {\scriptsize\textcolor{gray}{$\uparrow$1.55}}
& 84.56 {\scriptsize\textcolor{gray}{$\uparrow$1.49}}
& 84.44 {\scriptsize\textcolor{gray}{$\uparrow$1.37}}
& 84.60 {\scriptsize\textcolor{gray}{$\uparrow$1.53}}
& 84.35 {\scriptsize\textcolor{gray}{$\uparrow$1.28}} \\

Self Refine
& 80.91
& 83.51 {\scriptsize\textcolor{gray}{$\uparrow$2.60}}
& 82.29 {\scriptsize\textcolor{gray}{$\uparrow$1.38}}
& 82.40 {\scriptsize\textcolor{gray}{$\uparrow$1.49}}
& 82.14 {\scriptsize\textcolor{gray}{$\uparrow$1.23}}
& 82.53 {\scriptsize\textcolor{gray}{$\uparrow$1.62}}
& 82.06 {\scriptsize\textcolor{gray}{$\uparrow$1.15}} \\

\midrule
Average
& 82.87
& \textbf{85.34} {\scriptsize\textcolor{gray}{$\uparrow$\textbf{2.47}}}
& 84.71 {\scriptsize\textcolor{gray}{$\uparrow$1.83}}
& 84.66 {\scriptsize\textcolor{gray}{$\uparrow$1.79}}
& 84.63 {\scriptsize\textcolor{gray}{$\uparrow$1.76}}
& 84.82 {\scriptsize\textcolor{gray}{$\uparrow$1.95}}
& 84.55 {\scriptsize\textcolor{gray}{$\uparrow$1.68}} \\

\bottomrule
\end{tabular}
\label{tab-cot-retrieval}
\end{table}

In our experiments with a context length of 1k and 256 tokens, we used GPT-4o-mini. The maximum number of generated tokens, the line length of the sliding window, and the sliding step were set to 1024, 1, and 1, respectively. The maximum number of retrieved code snippets was set to 2. 

The improvement is more pronounced with a context length of 256 than with 1k. Among these, the LCS method stands out as the most practical, delivering top results for both short contexts and long contexts at minimal computational cost. As shown in Table~\ref{tab-cot-retrieval}, the performance of various code repair methods improved after introducing the retrieval mechanism, indicating that incorporating context (obtained through retrieval) can effectively enhance the performance of code repair. It can also be observed that the improvement with a context length of 256 is much greater than that with a context length of 1k. When the context length is set to 1k tokens, the average improvement brought by different retrieval methods was similar. The best-performing method is LCS, which provided an average improvement of approximately 2.5\%. When the context length is 256 tokens, the best-performing method is also LCS, and it clearly outperforms other retrieval methods. As a sparse retrieval method, LCS requires fewer computational resources compared to dense retrieval methods. Therefore, the LCS method is the most practical, achieving the best results with very low computational costs, regardless of whether the context length is small or large.
\begin{answer}
    \textbf{Finding 2:} Introducing a retrieval mechanism significantly improves the performance of various code repair methods, with LCS achieving the best results.
\end{answer}

\begin{table}[tb]
\caption{
Performance of Retrieval-Augmented Repair (RAR) Methods Using GPT-4o-mini with the best-performing retrieval method LCS. ``No Repair'' refers to single generation without RAR. ``Avg.'' shows the average performance across context lengths. ``Avg.\ RAR'' measures average performance at specific lengths, and ``Increase'' indicates improvement over ``No Repair''. P@1 and C@1 denote pass@1 and compilation@1.}
\centering
\scriptsize 
    \begin{tabular}{@{}lcccccccccccccc@{}}
        \toprule
        \textbf{\multirow{3}{*}{CR Method}} & \multicolumn{12}{c}{\textbf{Context Length}} & \multicolumn{2}{c}{} \\
        \cmidrule(lr){2-13}
         &     \multicolumn{2}{c}{\textbf{0}} &     \multicolumn{2}{c}{\textbf{256}} &     \multicolumn{2}{c}{\textbf{512}} &     \multicolumn{2}{c}{\textbf{1k}} &     \multicolumn{2}{c}{\textbf{2k}} &     \multicolumn{2}{c}{\textbf{4k}} & \multicolumn{2}{c}{\textbf{Avg.}} \\
         \cmidrule(lr){2-3} \cmidrule(lr){4-5} \cmidrule(lr){6-7}    \cmidrule(lr){8-9} \cmidrule(lr){10-11} \cmidrule(lr){12-13} \cmidrule(lr){14-15}
         & P@1 & C@1 & P@1 & C@1 & P@1 & C@1 & P@1 & C@1 & P@1 & C@1 & P@1 & C@1 & P@1 & C@1 \\
        \midrule
        No Repair & 38.88 & 38.95 & 66.10 & 66.20 & 72.89 & 73.01 & 80.19 & 80.33 & 85.93 & 86.21 & 88.87 & 89.20 & 72.14 & 72.32 \\
        Self Edit & \textbf{56.61} & 56.82 & \textbf{77.46} & 77.69 & \textbf{82.17} & 82.42 & \textbf{86.81} & 87.13 & \textbf{90.51} & 91.09 & \textbf{92.48} & 93.15 & \textbf{81.01} & 81.38 \\
        Self Debug & 53.04 & 53.24 & 75.04 & 75.23 & 80.08 & 80.27 & 85.50 & 85.74 & 89.56 & 90.00 & 91.64 & 92.12 & 79.14 & 79.43 \\
        Self Repair & 52.40 & 52.54 & 74.37 & 74.55 & 79.59 & 79.77 & 85.16 & 85.39 & 89.71 & 90.23 & 92.14 & 92.73 & 78.90 & 79.20 \\
        Self Refine & 48.77 & 48.92 & 71.92 & 72.11 & 77.54 & 77.74 & 83.38 & 83.65 & 87.90 & 88.43 & 90.40 & 91.02 & 76.65 & 76.98 \\
        \midrule
        Avg. RAR & 52.71 & 52.88 & 74.70 & 74.90 & 79.85 & 80.05 & 85.21 & 85.48 & 89.42 & 89.94 & 91.67 & 92.26 & 78.92 & 79.25 \\
        Increase (\%) & 13.83 & 13.93 & 8.60 & 8.70 & 6.96 & 7.04 & 5.02 & 5.15 & 3.49 & 3.73 & 2.79 & 3.05 & 6.78 & 6.93 \\
        \bottomrule
        \end{tabular}

\label{tab-retrieval-length}
\end{table}
        \paragraph{Assessing Performance Under Different Context Length}
        \label{sec-RAR-double}
We also tested the effectiveness of the Retrieval-Augmented Repair (RAR) framework under different context lengths. In our experiments, we used GPT-4o-mini with the Longest Common Subsequence (LCS) as the retrieval method. 

Table~\ref{tab-retrieval-length} demonstrates that RAR substantially improves both P@1 and C@1 across all context lengths. At context length 0, RAR yields an average improvement of 13.83\% in P@1 and 13.93\% in C@1, with \emph{Self Edit} achieving the best performance (56.61\% / 56.82\%). 
Remarkably, RAR allows models with shorter context lengths to match or exceed the performance of models with much longer contexts but without RAR. 
For instance, RAR enables models at 1k context to match the performance of no-RAR models at 2k context (85.21\% vs. 85.93\% in P@1), and \emph{Self Edit} further pushes this to 86.81\%. Similarly, at 2k context, \emph{Self Edit} even surpasses the no-RAR 4k baseline (90.51\% vs. 88.87\%).

This demonstrates that RAR can effectively compensate for limited context by providing relevant retrieved information, enabling the model to perform comparably to having much longer inputs. 
Even in scenarios where computation is less constrained, RAR continues to provide tangible benefits. At 4k context, where models already perform strongly, RAR still delivers an average improvement of 2.79\% in P@1 and 3.05\% in C@1.

Additionally, in the ``Avg.\ Ctx\ Len'' column, \emph{Self Refine}, which previously underperformed due to the lack of executor feedback, shows marked improvement under RAR, with over 4\% average gains in both metrics. This suggests that the retrieval mechanism itself provides valuable external guidance, reinforcing the utility of RAR beyond feedback-dependent methods.
\begin{answer}
    \textbf{Finding 3:} RAR improves code completion performance equivalent to doubling the context length. It also notably lifts methods like Self Refine by supplying useful external guidance.
\end{answer}

\subsubsection{RQ4: How effective is RAR in reducing inference costs?}
\label{sec-cost-effective}
\begin{figure}[tbp]
    \centering
    \includegraphics[width=0.39\textwidth]{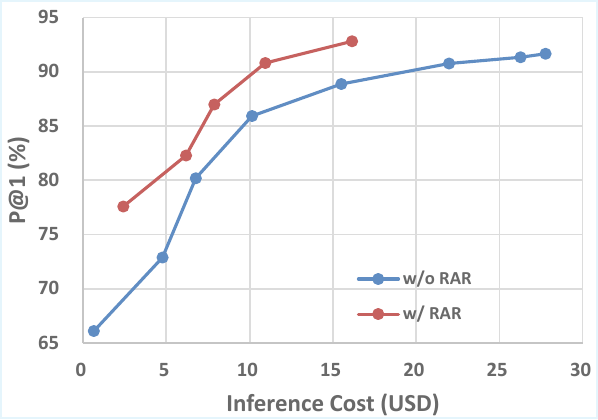} 
    \caption{Cost-effectiveness of RAR on SolBench with GPT-4o-mini. RAR uses LCS as retriever and Self Edit for repair.}
    \label{fig-cost-effective} 
\end{figure}
To assess the cost-effectiveness of the RAR framework, we measured the total cost for GPT-4o-mini on SolBench, both with and without RAR. For these experiments, RAR utilized the LCS as the retriever and Self Edit as the code repair method. The OpenAI API pricing for GPT-4o-mini is \$0.15 per million prompt tokens (input) and \$0.6 per million completion tokens (output). Figure~\ref{fig-cost-effective} shows P@1 vs. inference cost: w/o RAR has 8 data points (context from 0 to 32k) and w/ RAR has 5 data points (context from 256 to 4k), both increasing with inference cost.

For any given P@1, the RAR framework consistently requires a lower inference cost compared to the baseline without RAR.
For example, to achieve a P@1 of 90.81\%, RAR requires only \$10.94, whereas the baseline needs \$21.97 to reach a slightly lower P@1 of 90.76\%. This represents a cost reduction of approximately 50.20\%
At lower cost levels, RAR’s advantage is even more pronounced: with just \$2.41, RAR attains a P@1 of 77.59\%, while the baseline achieves only 72.89\% at a higher cost of \$4.77. In other words, RAR not only \textbf{halves} the inference cost but also improves accuracy by 5\%, underscoring its superior efficiency and effectiveness compared to the baseline.
\begin{answer}
    \textbf{Answer to RQ4:} RAR is highly cost-effective, consistently achieving equal or higher P@1 than the baseline at less than half the inference cost by operating with much shorter contexts.
\end{answer}

\section{Threats to Validity}
\subsection{External Validity}
While our evaluation and the RAR framework are implemented in Solidity on Ethereum, functional correctness verification relies on Solidity-specific tools (e.g., Diffusc and Echidna). These tools would need to be replaced with appropriate execution or testing frameworks when porting to other languages or platforms. In contrast, the core component of RAR operates on executor error messages and retrieval methods (e.g., LCS, BM25), which are language-agnostic and can potentially be adapted beyond Solidity.

\subsection{Internal Validity}
Some functions in SolBench may overlap with the training data of the evaluated LLMs, which could potentially inflate their performance. However, Solidity contracts exhibit a high function-level duplication rate (up to 86.85\%), indicating that code reuse is prevalent in real-world development scenarios. Therefore, even if such overlap exists, the performance of models on SolBench remains a strong indicator of their effectiveness in practical development settings. Additionally, differences in prompt construction, retrieval strategies, or inference settings could influence the performance of both the RAR framework and the baseline models. Therefore, the efficiency results should be interpreted in the context of our experimental setup.

\subsection{Construct Validity}
Our evaluation focuses on functional correctness and does not consider other important aspects such as code readability, maintainability, or gas efficiency, which are also critical in practical smart contract development. Future work could incorporate additional metrics to provide a more holistic assessment.
\section{Related Work}

\subsection{Evaluation of Code Generation}
Early evaluations of code LLMs rely on lexical overlap metrics such as BLEU and CrystalBLEU~\cite{CrystalBLEU,barone2017parallel}. While convenient, these metrics correlate weakly with functional correctness because many valid implementations differ in surface form (naming, control flow, statement ordering). In contrast, function-level, test-based evaluation has become standard in Python through benchmarks such as HumanEval~\cite{chen2021evaluating}, APPS~\cite{hendrycks2021measuring}, MBPP~\cite{austin2021program}, and DS-1000~\cite{lai2023ds}, which directly measure whether generated programs satisfy specifications. However, Solidity lacks a similarly large-scale, automated framework for function-level correctness. To addresses this gap, our work propose SolBench: a Solidity-specific benchmark and execution pipeline that uses differential fuzzing to verify functional equivalence, enabling robust and scalable correctness evaluation beyond BLEU-style proxies.

\subsection{Repair and Retrieval-Augmented Generation}
Automatic code repair techniques improve correctness by leveraging execution traces or feedback to iteratively fix code. Approaches such as trace-guided Self Edit~\cite{zhang2023selfeditfaultawarecodeeditor}, Self Debug~\cite{chen2023teachinglargelanguagemodels} and Self Correction~\cite{welleck2022generatingsequenceslearningselfcorrect} demonstrate that lightweight feedback loops can substantially raise pass rates. In parallel, retrieval-augmented code generation incorporates auxiliary code snippets or documentation retrieved from external repositories to ground generation and reduce hallucination~\cite{lewis2020rag, zhao2023verify-edit, gao2023ragsurvey, xu2024pds}. Our Retrieval-Augmented Repair differs in two ways: it (1) retrieves only from the provided intra-contract context rather than external corpora, and (2) uses executor feedback to guide retrieval and repair, allowing accurate completion under tight context windows while reducing inference cost.

\subsection{Solidity-specific Landscape: Datasets, Evaluation, and Tasks}
Large Solidity corpora have been collected from the Ethereum mainnet and Etherscan~\cite{storhaug2023efficient,yu2024smartllamatwostageposttraininglarge,morello2024dislfuelingresearchlarge}. Yet these resources are not directly suited to function-level, correctness-based evaluation. Prior assessments of Solidity code generation often rely on BLEU~\cite{zhang2025codebc}, CrystalBLEU~\cite{storhaug2023efficient} or manual grading~\cite{dade2023optimizing}, which do not robustly capture correctness. 

Recent work has begun to explore functional evaluation for Solidity~\cite{10732686}. Concurrent with our work, SolEval~\cite{peng2025solevalbenchmarkinglargelanguage} evaluates generated code using repository-provided or manually written tests on a dataset of 1,125 samples drawn from 9 repositories across 6 domains. However, SolEval’s reliance on such tests limits its scalability.
In contrast, SolBench offers a fully automated functional correctness pipeline based on differential fuzzing, requiring no human-written tests. It includes 28,825 functions from 7,604 real-world contracts spanning the full Etherscan history (genesis–2024) and 10 domains, substantially exceeding SolEval in both scale and diversity. Moreover, our work introduces a retrieval-augmented repair framework that improves correctness while reducing inference cost by approximately 50\%.

Beyond generation, LLMs have been studied for adjacent smart-contract tasks such as vulnerability detection~\cite{zhang2024acfix} and repair, fuzzing guidance~\cite{shou2024llm4fuzz}, summarization\cite{zhao2024automatic}, and gas-cost optimization~\cite{10757316}, underscoring the need for evaluation that respects Solidity’s execution semantics. 
Besides, prior work has examined the evolution and real-world usage of Solidity’s error-handling mechanisms~\cite{10.1145/3674805.3686686}. SmartBugs~\cite{10.1145/3324884.3415298} contributes an extensible execution framework that unifies multiple static and dynamic analysis tools and datasets for Solidity smart contracts.

\section{Conclusion}
In this work, we introduced SolBench, a large-scale benchmark and execution-based evaluation pipeline specifically designed for smart contracts written in Solidity. Unlike prior approaches that rely on surface-form similarity metrics such as BLEU, SolBench rigorously measures the functional correctness of LLM-generated smart contracts through differential fuzzing. Our results demonstrate that functional correctness is essential, as surface-level metrics poorly correlate with actual execution behavior.
Using SolBench, we conducted a comprehensive evaluation of 14 LLMs and found that Solidity code completion strongly depends on intra-contract context. While naively scaling context windows can partially alleviate this issue, it incurs substantial inference costs and yields diminishing returns.
To address this limitation, we proposed the Retrieval-Augmented Repair (RAR) framework, which leverages executor feedback to selectively retrieve the most relevant intra-contract code snippets and iteratively repair incorrect generations. Our results show that RAR consistently outperforms naïve long-context scaling, achieving higher functional correctness while reducing inference cost by approximately half.
Together, SolBench and RAR provide a foundation for evaluating and improving LLM-based smart contract generation, enabling cost-effective and reliable LLM-assisted smart contract development for both researchers and practitioners.

Looking ahead, we plan to continuously update SolBench to keep pace with the evolving Solidity ecosystem and compiler versions. This will help further narrow the gap between LLM-generated Solidity code and production-grade smart contracts.

\section{Data Availability}
The SolBench dataset and code are available at \url{https://github.com/ZaoyuChen/SolBench}.

\bibliographystyle{ACM-Reference-Format}
\bibliography{7anthology,7References}

@article{chen2021evaluating,
  title={Evaluating large language models trained on code},
  author={Chen, Mark and Tworek, Jerry and Jun, Heewoo and Yuan, Qiming and Pinto, Henrique Ponde de Oliveira and Kaplan, Jared and Edwards, Harri and Burda, Yuri and Joseph, Nicholas and Brockman, Greg and others},
  journal={arXiv preprint arXiv:2107.03374},
  year={2021}
}

@article{nijkamp2022codegen,
  title={Codegen: An open large language model for code with multi-turn program synthesis},
  author={Nijkamp, Erik and Pang, Bo and Hayashi, Hiroaki and Tu, Lifu and Wang, Huan and Zhou, Yingbo and Savarese, Silvio and Xiong, Caiming},
  journal={arXiv preprint arXiv:2203.13474},
  year={2022}
}

@article{li2023starcoder,
  title={Starcoder: may the source be with you!},
  author={Li, Raymond and Allal, Loubna Ben and Zi, Yangtian and Muennighoff, Niklas and Kocetkov, Denis and Mou, Chenghao and Marone, Marc and Akiki, Christopher and Li, Jia and Chim, Jenny and others},
  journal={arXiv preprint arXiv:2305.06161},
  year={2023}
}

@article{guo2024deepseek,
  title={DeepSeek-Coder: When the Large Language Model Meets Programming--The Rise of Code Intelligence},
  author={Guo, Daya and Zhu, Qihao and Yang, Dejian and Xie, Zhenda and Dong, Kai and Zhang, Wentao and Chen, Guanting and Bi, Xiao and Wu, Y and Li, YK and others},
  journal={arXiv preprint arXiv:2401.14196},
  year={2024}
}

@inproceedings{zhang2023repocoder,
  title={RepoCoder: Repository-Level Code Completion Through Iterative Retrieval and Generation},
  author={Zhang, Fengji and Chen, Bei and Zhang, Yue and Keung, Jacky and Liu, Jin and Zan, Daoguang and Mao, Yi and Lou, Jian-Guang and Chen, Weizhu},
  booktitle={Proceedings of the 2023 Conference on Empirical Methods in Natural Language Processing},
  pages={2471--2484},
  year={2023}
}

@inproceedings{shrivastava2023repository,
  title={Repository-level prompt generation for large language models of code},
  author={Shrivastava, Disha and Larochelle, Hugo and Tarlow, Daniel},
  booktitle={International Conference on Machine Learning},
  pages={31693--31715},
  year={2023},
  organization={PMLR}
}

@article{dade2023optimizing,
  title={Optimizing Large Language Models to Expedite the Development of Smart Contracts},
  author={Dade, Nii Osae Osae and Lartey-Quaye, Margaret and Odonkor, Emmanuel Teye-Kofi and Ammah, Paul},
  journal={arXiv preprint arXiv:2310.05178},
  year={2023}
}

@article{roziere2023code,
  title={Code llama: Open foundation models for code},
  author={Roziere, Baptiste and Gehring, Jonas and Gloeckle, Fabian and Sootla, Sten and Gat, Itai and Tan, Xiaoqing Ellen and Adi, Yossi and Liu, Jingyu and Remez, Tal and Rapin, J{\'e}r{\'e}my and others},
  journal={arXiv preprint arXiv:2308.12950},
  year={2023}
}

@article{hendrycks2021measuring,
  title={Measuring coding challenge competence with apps},
  author={Hendrycks, Dan and Basart, Steven and Kadavath, Saurav and Mazeika, Mantas and Arora, Akul and Guo, Ethan and Burns, Collin and Puranik, Samir and He, Horace and Song, Dawn and others},
  journal={arXiv preprint arXiv:2105.09938},
  year={2021}
}

@article{austin2021program,
  title={Program synthesis with large language models},
  author={Austin, Jacob and Odena, Augustus and Nye, Maxwell and Bosma, Maarten and Michalewski, Henryk and Dohan, David and Jiang, Ellen and Cai, Carrie and Terry, Michael and Le, Quoc and others},
  journal={arXiv preprint arXiv:2108.07732},
  year={2021}
}

@inproceedings{storhaug2023efficient,
  title={Efficient avoidance of vulnerabilities in auto-completed smart contract code using vulnerability-constrained decoding},
  author={Storhaug, Andr{\'e} and Li, Jingyue and Hu, Tianyuan},
  booktitle={2023 IEEE 34th International Symposium on Software Reliability Engineering (ISSRE)},
  pages={683--693},
  year={2023},
  organization={IEEE}
}

@article{liu2023repobench,
  title={Repobench: Benchmarking repository-level code auto-completion systems},
  author={Liu, Tianyang and Xu, Canwen and McAuley, Julian},
  journal={arXiv preprint arXiv:2306.03091},
  year={2023}
}

@inproceedings{Feist_2019_Slither,
   title={Slither: A Static Analysis Framework for Smart Contracts},
   url={http://dx.doi.org/10.1109/WETSEB.2019.00008},
   DOI={10.1109/wetseb.2019.00008},
   booktitle={2019 IEEE/ACM 2nd International Workshop on Emerging Trends in Software Engineering for Blockchain (WETSEB)},
   publisher={IEEE},
   author={Feist, Josselin and Grieco, Gustavo and Groce, Alex},
   year={2019},
   month=may }

@inproceedings{Grieco2020Echidna,
author = {Grieco, Gustavo and Song, Will and Cygan, Artur and Feist, Josselin and Groce, Alex},
title = {Echidna: effective, usable, and fast fuzzing for smart contracts},
year = {2020},
isbn = {9781450380089},
publisher = {Association for Computing Machinery},
address = {New York, NY, USA},
url = {https://doi.org/10.1145/3395363.3404366},
doi = {10.1145/3395363.3404366},
booktitle = {Proceedings of the 29th ACM SIGSOFT International Symposium on Software Testing and Analysis},
pages = {557–560},
numpages = {4},
location = {Virtual Event, USA},
series = {ISSTA 2020}
}

@misc{CryticDiffusc2024,
  author = {Crytic},
  title = {Crytic/Diffusc},
  howpublished = {GitHub},
  year = {2024},
  note = {Accessed: 2024-06-13, \url{https://github.com/crytic/diffusc}}
}

@article{feng2020codebert,
  title={Codebert: A pre-trained model for programming and natural languages},
  author={Feng, Zhangyin and Guo, Daya and Tang, Duyu and Duan, Nan and Feng, Xiaocheng and Gong, Ming and Shou, Linjun and Qin, Bing and Liu, Ting and Jiang, Daxin and others},
  journal={arXiv preprint arXiv:2002.08155},
  year={2020}
}

@article{guo2022unixcoder,
  title={Unixcoder: Unified cross-modal pre-training for code representation},
  author={Guo, Daya and Lu, Shuai and Duan, Nan and Wang, Yanlin and Zhou, Ming and Yin, Jian},
  journal={arXiv preprint arXiv:2203.03850},
  year={2022}
}

@inproceedings{lai2023ds,
  title={DS-1000: A natural and reliable benchmark for data science code generation},
  author={Lai, Yuhang and Li, Chengxi and Wang, Yiming and Zhang, Tianyi and Zhong, Ruiqi and Zettlemoyer, Luke and Yih, Wen-tau and Fried, Daniel and Wang, Sida and Yu, Tao},
  booktitle={International Conference on Machine Learning},
  pages={18319--18345},
  year={2023},
  organization={PMLR}
}

@article{zhao2024automatic,
  title={Automatic smart contract comment generation via large language models and in-context learning},
  author={Zhao, Junjie and Chen, Xiang and Yang, Guang and Shen, Yiheng},
  journal={Information and Software Technology},
  volume={168},
  pages={107405},
  year={2024},
  publisher={Elsevier}
}

@article{zhang2024acfix,
  title={Acfix: Guiding llms with mined common rbac practices for context-aware repair of access control vulnerabilities in smart contracts},
  author={Zhang, Lyuye and Li, Kaixuan and Sun, Kairan and Wu, Daoyuan and Liu, Ye and Tian, Haoye and Liu, Yang},
  journal={arXiv preprint arXiv:2403.06838},
  year={2024}
}

@article{shou2024llm4fuzz,
  title={LLM4Fuzz: Guided Fuzzing of Smart Contracts with Large Language Models},
  author={Shou, Chaofan and Liu, Jing and Lu, Doudou and Sen, Koushik},
  journal={arXiv preprint arXiv:2401.11108},
  year={2024}
}

@article{ding2024crosscodeeval,
  title={Crosscodeeval: A diverse and multilingual benchmark for cross-file code completion},
  author={Ding, Yangruibo and Wang, Zijian and Ahmad, Wasi and Ding, Hantian and Tan, Ming and Jain, Nihal and Ramanathan, Murali Krishna and Nallapati, Ramesh and Bhatia, Parminder and Roth, Dan and others},
  journal={Advances in Neural Information Processing Systems},
  volume={36},
  year={2024}
}

@article{lewis2020rag,
  title={Retrieval-augmented generation for knowledge-intensive nlp tasks},
  author={Lewis, Patrick and Perez, Ethan and Piktus, Aleksandra and Petroni, Fabio and Karpukhin, Vladimir and Goyal, Naman and K{\"u}ttler, Heinrich and Lewis, Mike and Yih, Wen-tau and Rockt{\"a}schel, Tim and others},
  journal={Advances in neural information processing systems},
  volume={33},
  pages={9459--9474},
  year={2020}
}

@article{zhao2023verify-edit,
  title={Verify-and-edit: A knowledge-enhanced chain-of-thought framework},
  author={Zhao, Ruochen and Li, Xingxuan and Joty, Shafiq and Qin, Chengwei and Bing, Lidong},
  journal={arXiv preprint arXiv:2305.03268},
  year={2023}
}

@article{gao2023ragsurvey,
  title={Retrieval-augmented generation for large language models: A survey},
  author={Gao, Yunfan and Xiong, Yun and Gao, Xinyu and Jia, Kangxiang and Pan, Jinliu and Bi, Yuxi and Dai, Yi and Sun, Jiawei and Wang, Haofen and Wang, Haofen},
  journal={arXiv preprint arXiv:2312.10997},
  volume={2},
  year={2023}
}

@article{xu2024pds,
  title={Can We Verify Step by Step for Incorrect Answer Detection?},
  author={Xu, Xin and Diao, Shizhe and Yang, Can and Wang, Yang},
  journal={arXiv preprint arXiv:2402.10528},
  year={2024}
}

@misc{zhang2023selfeditfaultawarecodeeditor,
      title={Self-Edit: Fault-Aware Code Editor for Code Generation}, 
      author={Kechi Zhang and Zhuo Li and Jia Li and Ge Li and Zhi Jin},
      year={2023},
      eprint={2305.04087},
      archivePrefix={arXiv},
      primaryClass={cs.SE},
      url={https://arxiv.org/abs/2305.04087}, 
}

@misc{chen2023teachinglargelanguagemodels,
      title={Teaching Large Language Models to Self-Debug}, 
      author={Xinyun Chen and Maxwell Lin and Nathanael Schärli and Denny Zhou},
      year={2023},
      eprint={2304.05128},
      archivePrefix={arXiv},
      primaryClass={cs.CL},
      url={https://arxiv.org/abs/2304.05128}, 
}

@misc{welleck2022generatingsequenceslearningselfcorrect,
      title={Generating Sequences by Learning to Self-Correct}, 
      author={Sean Welleck and Ximing Lu and Peter West and Faeze Brahman and Tianxiao Shen and Daniel Khashabi and Yejin Choi},
      year={2022},
      eprint={2211.00053},
      archivePrefix={arXiv},
      primaryClass={cs.CL},
      url={https://arxiv.org/abs/2211.00053}, 
}

@misc{yu2024smartllamatwostageposttraininglarge,
      title={Smart-LLaMA: Two-Stage Post-Training of Large Language Models for Smart Contract Vulnerability Detection and Explanation}, 
      author={Lei Yu and Shiqi Chen and Hang Yuan and Peng Wang and Zhirong Huang and Jingyuan Zhang and Chenjie Shen and Fengjun Zhang and Li Yang and Jiajia Ma},
      year={2024},
      eprint={2411.06221},
      archivePrefix={arXiv},
      primaryClass={cs.CR},
      url={https://arxiv.org/abs/2411.06221}, 
}

@misc{morello2024dislfuelingresearchlarge,
      title={DISL: Fueling Research with A Large Dataset of Solidity Smart Contracts}, 
      author={Gabriele Morello and Mojtaba Eshghie and Sofia Bobadilla and Martin Monperrus},
      year={2024},
      eprint={2403.16861},
      archivePrefix={arXiv},
      primaryClass={cs.SE},
      url={https://arxiv.org/abs/2403.16861}, 
}

@misc{madaan2023selfrefineiterativerefinementselffeedback,
      title={Self-Refine: Iterative Refinement with Self-Feedback}, 
      author={Aman Madaan and Niket Tandon and Prakhar Gupta and Skyler Hallinan and Luyu Gao and Sarah Wiegreffe and Uri Alon and Nouha Dziri and Shrimai Prabhumoye and Yiming Yang and Shashank Gupta and Bodhisattwa Prasad Majumder and Katherine Hermann and Sean Welleck and Amir Yazdanbakhsh and Peter Clark},
      year={2023},
      eprint={2303.17651},
      archivePrefix={arXiv},
      primaryClass={cs.CL},
      url={https://arxiv.org/abs/2303.17651}, 
}

@misc{olausson2024selfrepairsilverbulletcode,
      title={Is Self-Repair a Silver Bullet for Code Generation?}, 
      author={Theo X. Olausson and Jeevana Priya Inala and Chenglong Wang and Jianfeng Gao and Armando Solar-Lezama},
      year={2024},
      eprint={2306.09896},
      archivePrefix={arXiv},
      primaryClass={cs.CL},
      url={https://arxiv.org/abs/2306.09896}, 
}

@misc{li2024codesbuildingopensourcelanguage,
      title={CodeS: Towards Building Open-source Language Models for Text-to-SQL}, 
      author={Haoyang Li and Jing Zhang and Hanbing Liu and Ju Fan and Xiaokang Zhang and Jun Zhu and Renjie Wei and Hongyan Pan and Cuiping Li and Hong Chen},
      year={2024},
      eprint={2402.16347},
      archivePrefix={arXiv},
      primaryClass={cs.CL},
      url={https://arxiv.org/abs/2402.16347}, 
}

@ARTICLE{10757316,
  author={Jiang, Jinan and Li, Zihao and Qin, Haoran and Jiang, Muhui and Luo, Xiapu and Wu, Xiaoming and Wang, Haoyu and Tang, Yutian and Qian, Chenxiong and Chen, Ting},
  journal={IEEE Transactions on Software Engineering}, 
  title={Unearthing Gas-Wasting Code Smells in Smart Contracts with Large Language Models}, 
  year={2024},
  volume={},
  number={},
  pages={1-26},
  doi={10.1109/TSE.2024.3491578}}

@article{morello2024disl,
  title={DISL: Fueling Research with A Large Dataset of Solidity Smart Contracts},
  author={Morello, Gabriele and Eshghie, Mojtaba and Bobadilla, Sofia and Monperrus, Martin},
  journal={arXiv preprint arXiv:2403.16861},
  year={2024}
}

@article{zhang2025codebc,
  title={CodeBC: A More Secure Large Language Model for Smart Contract Code Generation in Blockchain},
  author={Zhang, Hainan and Zhang, Qinnan and Wang, Ziwei and Zheng, Hongwei and Dong, Jin and Zheng, Zhiming and others},
  journal={arXiv preprint arXiv:2504.21043},
  year={2025}
}

@unpublished{peng2025solevalbenchmarkinglargelanguage,
      title={SolEval: Benchmarking Large Language Models for Repository-level Solidity Code Generation}, 
      author={Zhiyuan Peng and Xin Yin and Rui Qian and Peiqin Lin and Yongkang Liu and Chenhao Ying and Yuan Luo},
      note={unpublished},
      year={2025},
      eprint={2502.18793},
      archivePrefix={arXiv},
      primaryClass={cs.SE},
      url={https://arxiv.org/abs/2502.18793}, 
}

@INPROCEEDINGS{10732686,
  author={Daspe, Etienne and Durand, Mathis and Hatin, Julien and Bradai, Salma},
  booktitle={2024 6th Conference on Blockchain Research \& Applications for Innovative Networks and Services (BRAINS)}, 
  title={Benchmarking Large Language Models for Ethereum Smart Contract Development}, 
  year={2024},
  volume={},
  number={},
  pages={1-4},
  doi={10.1109/BRAINS63024.2024.10732686}}

@article{barone2017parallel,
  title={A parallel corpus of python functions and documentation strings for automated code documentation and code generation},
  author={Barone, Antonio Valerio Miceli and Sennrich, Rico},
  journal={arXiv preprint arXiv:1707.02275},
  year={2017}
}

@inproceedings{CrystalBLEU,
author = {Eghbali, Aryaz and Pradel, Michael},
title = {CrystalBLEU: Precisely and Efficiently Measuring the Similarity of Code},
year = {2023},
isbn = {9781450394758},
publisher = {Association for Computing Machinery},
address = {New York, NY, USA},
url = {https://doi.org/10.1145/3551349.3556903},
doi = {10.1145/3551349.3556903},
booktitle = {Proceedings of the 37th IEEE/ACM International Conference on Automated Software Engineering},
articleno = {28},
numpages = {12},
location = {Rochester, MI, USA},
series = {ASE '22}
}

@article{Kocetkov2022TheStack,
  title={The Stack: 3 TB of permissively licensed source code},
  author={Kocetkov, Denis and Li, Raymond and Ben Allal, Loubna and Li, Jia and Mou,Chenghao and Muñoz Ferrandis, Carlos and Jernite, Yacine and Mitchell, Margaret and Hughes, Sean and Wolf, Thomas and Bahdanau, Dzmitry and von Werra, Leandro and de Vries, Harm},
  journal={Preprint},
  year={2022}
}

@inproceedings{10.1145/3674805.3686686,
author = {Mitropoulos, Charalambos and Kechagia, Maria and Maschas, Chrysostomos and Ioannidis, Sotirios and Sarro, Federica and Mitropoulos, Dimitrios},
title = {Broken Agreement: The Evolution of Solidity Error Handling},
year = {2024},
isbn = {9798400710476},
publisher = {Association for Computing Machinery},
address = {New York, NY, USA},
url = {https://doi.org/10.1145/3674805.3686686},
doi = {10.1145/3674805.3686686},
booktitle = {Proceedings of the 18th ACM/IEEE International Symposium on Empirical Software Engineering and Measurement},
pages = {257–268},
numpages = {12},
location = {Barcelona, Spain},
series = {ESEM '24}
}

@inproceedings{10.1145/3324884.3415298,
author = {Ferreira, Jo\~{a}o F. and Cruz, Pedro and Durieux, Thomas and Abreu, Rui},
title = {SmartBugs: a framework to analyze solidity smart contracts},
year = {2021},
isbn = {9781450367684},
publisher = {Association for Computing Machinery},
address = {New York, NY, USA},
url = {https://doi.org/10.1145/3324884.3415298},
doi = {10.1145/3324884.3415298},
booktitle = {Proceedings of the 35th IEEE/ACM International Conference on Automated Software Engineering},
pages = {1349–1352},
numpages = {4},
location = {Virtual Event, Australia},
series = {ASE '20}
}

\end{document}